\documentclass[12pt]{iopart}
\usepackage{iopams}
\def\Quadrat#1#2{{\vcenter{\hrule height #2
\hbox{\vrule width #2 height #1 \kern#1
\vrule width #2}\hrule height #2}}}
\def\Box{\mathop{\kern 1pt\hbox{$\Quadrat{8pt}{0.4pt}$} \kern 1pt}}
\usepackage{amssymb}
\usepackage[dvips]{graphics}
\usepackage{epsfig}
\usepackage{bm}
\expandafter\ifx\csname package@font\endcsname\relax\else
 \expandafter\expandafter
 \expandafter\usepackage
 \expandafter\expandafter
 \expandafter{\csname package@font\endcsname}%
\fi
\newcommand{\SII}{section 2\;}
\newcommand{\SIII}{section 3\;}
\newcommand{\SIV}{section 4\;}
\newcommand{\SV}{section 5\;}
\newcommand{\SVI}{section 6\;}
\newcommand{\aj}{{\it Astron. J. (USA)}}

\begin{document}
\title[Gravimagnetism in the Gravitational Light-Ray Deflection Experiments]{Gravimagnetism, Causality, and Aberration of Gravity in the Gravitational Light-Ray Deflection Experiments}
\author{Sergei M. Kopeikin} 
\address{Department of Physics \& Astronomy, University of
Missouri, Columbia, Missouri 65211}
\ead{kopeikins@missouri.edu}
\author{Edward B. Fomalont}
\address{National Radio Astronomy Observatory, Charlottesville, VA 
22903, USA}
\ead{efomalon@nrao.edu}
\date{\today}
\begin{abstract}\noindent
Experimental verification of the existence of gravimagnetic fields
generated by currents of matter is important for a complete
understanding and formulation of gravitational physics.  Although the
rotational ({\it intrinsic}) gravimagnetic field has been
extensively studied and is now being measured by the Gravity Probe B,
the {\it extrinsic} gravimagnetic field generated by the
translational current of matter is less well studied.  The present
paper 
uses the post-Newtonian parametrized Einstein and light geodesics equations to show
that the {\it extrinsic} gravimagnetic field generated by the
translational current of matter can be measured by observing the
relativistic time delay and/or light deflection caused by the moving mass.  We prove that
the {\it extrinsic} gravimagnetic field is generated by the
relativistic effect of the aberration of the gravity force caused by
the Lorentz transformation of the metric tensor and the Levi-Civita
connection. We show that the Lorentz transformation of the gravity field 
variables is equivalent to the technique of the retarded Lienard-Wiechert 
gravitational potentials predicting that a light particle is deflected by 
gravitational field of a moving body from its retarded position so that 
both general-relativistic phenomena -- the aberration and the retardation of gravity --
are tightly connected and observing the aberration of gravity proves that gravity has a causal nature. We explain in this framework the 2002 deflection
experiment of a quasar by Jupiter where the aberration of gravity from
its orbital motion was measured with accuracy 20\%.  We describe a theory of VLBI
experiment to measure the gravitational deflection of radio waves
from a quasar by the Sun, as viewed by a moving observer from the geocentric frame, to improve the measurement accuracy of the aberration of gravity to a few percent.
\end{abstract}
\maketitle
\newpage
\newpage
\section{Introduction}

A gravimagnetic field, according to Einstein's theory of general
relativity, arises from moving matter (mass current) just as the
magnetic field arises in Maxwell's theory from moving charge
(electric current). The weak-field linearized theory of general
relativity unveils a mathematical structure comparable to the Maxwell
equations \cite{bct,m1,rugt}.  Hence, this weak-field approximation splits
gravitation into components similar to the electric and magnetic
field. In the case of the gravitational field, the source is the mass
of the body, whereas in the case of the electromagnetic field, the
source is the charge of the particle.  Moving the charge particle
creates a magnetic field according to Amp\`ere's law. Analogously,
moving the mass creates a mass current which generates a
gravimagnetic field according to Einstein's general relativity.
Amp\`ere-like induction of a gravimagnetic field (gravimagnetic
induction) in general relativity has been a matter of theoretical
study since the Lense-Thirring paper \cite{lt,cw,lneo}. Now, this
problem can be tackled experimentally.

There are two types of mass currents in gravity \cite{kop-ijmpd}. The first type is
produced by the intrinsic rotation of matter around body's center of
mass. It generates an {\it intrinsic} gravimagnetic field tightly
associated with body's angular momentum (spin) and most research in
gravimagnetism has been focused on the discussion of its various
properties \cite{cw}. Our recent publications \cite{kop-ijmpd,km,ckmr,wkop} as well as papers by other researchers \cite{kn}--\cite{1-12} give a comprehensive review of various aspects of the {\it intrinsic} gravimagnetism. It is interesting to note that the {\it intrinsic} gravimagnetic field can be associated with the holonomy invariance group \cite{mmm}. Some authors \cite{cam1,cam2} have proposed to measure the {\it intrinsic} gravimagnetic field by observing quantum effects of coupling of fermion's spin with the angular momentum of the Earth. It might be worthwhile to explore association of the {\it intrinsic} gravimagnetism with the classic Hannay precession phase \cite{hannay} \footnote[1]{This question is intriguing but nothing has been done to clarify this issue except of some papers by Spallicci \cite{spal1,spal2}.}.   

The first classic experiment to test the {\it intrinsic} gravimagnetic effect of the rotating Earth has been carried out by observing LAGEOS in combination with other geodetic satellites \cite{ciuf,cp} (see also \cite{cp-1,cp-2,cp-3}) which verified its existence as predicted by
Einstein's general relativity. Independent experimental measurement of the {\it intrinsic} gravimagnetic field of the rotating Earth is currently under way by
the Gravity Probe B mission \cite{gpb,cwil} that is expected to increment the accuracy of the preceding LAGEOS measurement.

The second type of the mass
current is caused by translational motion of matter. It generates an
{\it extrinsic} gravimagnetic field that depends on the frame of
reference of observer and can be completely eliminated in the rest
frame of the matter. This property of the {\it extrinsic} gravimagnetic
field is a direct consequence of the Lorentz invariance of Einstein's
gravity field equations for an isolated astronomical system \cite{fock-1,fock-2} embedded to the asymptotically-flat space-time and its experimental testing is as important
as that of the {\it intrinsic} gravimagnetic field. The point is
that both the {\it intrinsic} and the {\it extrinsic} gravimagnetic
fields obey the same equations and, therefore, their measurements
essentially complement each other \cite{kop-ijmpd}. Furthermore, detection of the {\it
extrinsic} gravimagnetic field probes the Lorentz invariance of the
gravitational field which determines the gravity null cone (domain of causal influence) on which the gravity force propagates. Experimental verification of the Lorentz invariance (causality) of gravity is important for the theory of braneworlds \cite{bbv} and for setting other, more stringent limitations on vector-tensor theories of gravity \cite{vtni}.

Ciufolini \cite{ciu} proposed to distinguish the {\it intrinsic} and {\it extrinsic} gravimagnetic fields by making use of two scalar invariants of the curvature tensor 
\begin{eqnarray}
\label{krin1}
I_1&=&R_{\alpha\beta\mu\nu}R^{\alpha\beta\mu\nu}\;,\\
\label{krin2}
I_2&=&R_{\alpha\beta}{}^{\mu\nu}R^{\alpha\beta\rho\sigma}E_{\mu\nu\rho\sigma}\;,
\end{eqnarray}
where $R_{\alpha\beta\mu\nu}$ is the curvature tensor, $E_{\mu\nu\rho\sigma}$ is the fully anti-symmetric Levi-Civita tensor with $E_{0123}=+\sqrt{-g}$, and $g={\rm det}(g_{\mu\nu})<0$ is the determinant of the metric tensor. 
Ciufolini \cite{ciu} notices that weak gravitational field of an isolated astronomical system yields $I_2=0$ if the {\it intrinsic} gravimagnetic field is absent. However, making use of $I_2$ is just one of many possibilities to single out the {\it intrinsic} gravimagnetic field. This question deserves further, more detailed study probably in the spirit of Petrov's algebraic classification \cite{petr} of gravitational fields or by making use of other techniques (see, for example, \cite{mmcq,prs}). One should not confuse the invariant $I_2$ with the gravimagnetic field itself. The gravimagnetic field is generated by {\it any} current of matter. Hence, $I_2=0$ does not mean that any gravimagnetic field is absent as assumed by some researchers \cite{pask}. Equality $I_2=0$ only implies that the gravimagnetic field is of the {\it extrinsic} origin ($I_1\not=0$), that is generated by a translational motion of matter. 

The goal of the present paper is to show that the {\it extrinsic}
gravimagnetic field can be measured in high-precision relativistic
time-delay experiments conducted in the Solar system where light (photon, radio wave) interacts with a
moving gravitational field of a massive body \cite{kopmak}.
The relativistic light deflection in the rest (static) frame of the
light-ray deflecting body is well-known and was calculated in optics
by Einstein \cite{ein}. Shapiro \cite{shapiro} has derived the
relativistic time delay for radio waves, and the general relativistic
problem of the gravitational deflection and delay of light by an
arbitrary moving point-like and spinning mass has been solved in our papers \cite{km,ks}.
Light-ray deflection experiments conducted in the gravitational field of a moving body allow us to study the global Lorentz transformation properties of the gravitational field since the angle of the gravitational light deflection is essentially an integral phenomenon that is not reduced to a local experiment in a particular laboratory frame where the effect of the gravitational field is reduced to the tidal force. Because both gravity and light participate in the gravitational deflection of light, one has to develop a formalism that is capable to trace the Lorentz transformation of the gravitational field variables and to distinguish it from that for light. The problem is that conventionally, in general relativity the speed of gravity is denoted by the same letter as the speed of light making an erroneous impression that gravity is associated with electromagnetism. The fundamental speed $c$ indeed enters many fundamental equations of modern physics but it has different physical facets which should be clearly distinguished to avoid confusion in the interpretation of physical laws observed experimentally \cite{ell}. 

To separate relativistic effects produced by the Lorentz transformation of gravitational field from those caused by the Lorentz transformation of electromagnetic field, we introduce in \SII a post-Newtonian parameter $\epsilon$ and put it in front of {\it all} time derivatives of the gravity field
variables of Einstein's equations by making use of their re-scaling 
\begin{equation}
\label{jtd}
{\bm j}\rightarrow\epsilon{\bm j}\;,\qquad\qquad{\bm
v}\rightarrow\epsilon{\bm v}\;,\qquad\qquad\frac{\partial }{\partial
t}\rightarrow\epsilon\frac{\partial }{\partial t}\;,
\end{equation}
where ${\bm j}$ and ${\bm v}$ are the matter current and velocity
respectively. 
Parameter $\epsilon$ can be interpreted as a label which goal is to track down the possible deviation of the null cone in general relativity from that in electrodynamics that can be characterized as the ratio $\epsilon\equiv c/c_{\rm g}$ with $c$ being the constant speed of light, and $c_{\rm g}$ taking a range of values
from $c_{\rm g}=\infty$ (Newtonian-like gravity) to $c_{\rm g}=c$ (general
relativity) \cite{fs,cqg,pla}. 
This parametrization of general relativity preserves the Newtonian
limit ($\epsilon=0$) of the Einstein equations and assumes that gravitational potentials and forces obey (in the background flat space-time) the Lorentz transformation that is parametrized by the speed of gravity $c_{\rm g}$. If $\epsilon$ were different from unity, the relativistic Lorentz transformation for gravity could be different from that in electrodynamics. This difference can be tested in the light-ray deflection experiments conducted in the field of a moving gravitating body because in such kind of experiments light interacts with time-dependent gravitational field and, hence, both speeds of interaction, $c$ and $c_{\rm g}$, are involved. The plausible difference of the speed of gravity, $c_{\rm g}$, from the speed of light, $c$, might be an indicator that the gravity force violates the principle of causality, that is propagate information faster than light. We demonstrate how to test this principle in the high-precision light-ray deflection experiments in the time-dependent field of a moving body.

In \SIII we show that the parameter $\epsilon$ is indeed a marker of the
Lorentz group of transformation of the gravitational field variables,
given by the matrix $\Lambda^\alpha_\beta(\epsilon)$ that depends on
the speed $c_{\rm g}\equiv c/\epsilon$.  This speed
defines the ultimate speed of propagation of the gravitational field in
Minkowski background space-time. The Lorentz group of the
electromagnetic field is described by the matrix
$\lambda^\alpha_\beta\equiv\Lambda^\alpha_\beta(\epsilon=1)$, and
depends on the speed of light $c$ in vacuum, so that the effects
associated with the null characteristics of the electromagnetic and
gravitational fields can be clearly discerned by measuring the numerical value of the parameter $\epsilon$.  

In \SIV we analyze the propagation of light, using the formalism from
the previous sections.  Utilizing this formalism, we show that relativistic equations associated with the
propagation of light distinguish between
$c_{\rm g}$ and $c$ already in linear terms of the order of $v/c{\rm_g}$ and $v/c$ beyond the static
Shapiro time delay.  The linear $v/c{\rm_g}$ correction is explained by the
relativistic effect of the aberration of gravity force associated with the gravimagnetic property of the
gravitational force propagating with the fundamental speed $c_{\rm g}$ from
the moving light-ray deflecting body to the light particle (photon) \cite{kfom}.

Although Einstein assumed that the Lorentz groups for gravitational and
electromagnetic field are identical \cite{LL,mtw}, this
theoretical prediction should be tested experimentally.  The
compatibility of the two groups can be measured by an experiment in
which both gravitational and electromagnetic fields are observed and transformed
simultaneously from one inertial frame to another. If the experiment is sufficiently
accurate, then the difference between the two
Lorentz-transformation speeds, $c$ and $c_{\rm g}$, can be measured.  Very Long Baseline
Interferometric (VLBI) measurements by observing the propagation of
radio waves (light) in the gravitational field of a moving massive
body does have sufficient sensitivity.  In \SV, we discuss the
physical interpretations of the deflection experiments similar to the
VLBI experiment in 2002 September during which Jupiter passed within
$3.7'$ of a bright quasar and the gravity-aberration term ($50~\mu$arcsec) was
measured to about 20\% accuracy ($10~\mu$arcsec) via retarded spatial coordinates of Jupiter in the gravitational time delay as observed in the barycentric frame \cite{k1,fk}.
A potentially more accurate experiment to measure the aberration (causality) of
gravity can be done when the Sun occults a
bright radio source like 3C279, and in \SVI we discuss theoretical details of the experiment.  

\section{Linearized Gravity and Gravimagnetism} 

Our $c_{\rm g}$-parametrized approach allows us to work out successive
approximations to solve the parametrized Einstein gravity field equations by
iterations with respect to various small parameters.  We shall use the
linearized approximation of the Einstein equations of
general relativity, expanded with respect to the universal
gravitational constant $G$. The linearized approximation is
currently sufficient for analysis of gravitational experiments within the solar
system. Terms of the quadratic order in $G$ may be detected in some future missions \cite{dittus}.  
When discussing the astrometric applications, the additional post-Newtonian expansion will be done with respect to the
parameter characterizing the ratio of spatial to temporal derivatives
of the metric tensor.  For slowly moving bodies and test particles this ratio is
\begin{equation}
\label{tq5}
\left|\frac{\partial/\partial x^0}{\partial/\partial
x^i}\right|=\epsilon\frac{v}{c}=\beta_\epsilon\;,
\end{equation}  
while for light-ray particles (for which $v=c$)
\begin{equation}
\label{tq7}
\left|\frac{\partial/\partial x^0}{\partial/\partial x^i}\right|=\epsilon\;,
\end{equation}  
where parameter $\epsilon=c/c_{\rm g}$. If $c_{\rm g}\rightarrow\infty$ (that is
$\epsilon\rightarrow 0$), the general relativity collapses to the
Newtonian-like theory in which gravitational waves have infinite speed so that the aberration of gravity force is absent and all other gravimagnetic phenomena also vanish
because the gravitational interaction is instantaneous. We retain in the post-Newtonian expansion of observable astrometric quantities all terms up to the order of $\epsilon^3$, and neglect terms of the higher order since they are not detectable with the current astrometric accuracy. 
 
In the linearized gravity the metric tensor perturbation
$h_{\alpha\beta}=g_{\alpha\beta}-\eta_{\alpha\beta}$ of the
gravitational field contains only the gravielectric,
$h_{00}=(2/c^2)\Phi$, and the gravimagnetic, $h_{0i}=-(4/c^2)A_i$,
potentials \cite{bct,m1,cw}. The space-space components of the metric
tensor perturbation $h_{ij}=(2/c^2)\Phi\delta_{ij}$, and the higher order terms are neglected. In the approximation under consideration the
gravielectric, ${\bm E}$, and the gravimagnetic, ${\bm B}$, fields
are formally defined as \cite{bct,m1,cw}
\begin{eqnarray}  
\label{1}
{\bm E}&=&-{\bm\nabla}\Phi-\frac{\epsilon}{c}\frac{\partial{\bm
A}}{\partial t}\;,\\
\label{2}
{\bm B}&=&{\bm\nabla}\times{\bm A}\;,
\end{eqnarray}
where ${\bm\nabla}\equiv\partial/\partial_i$ is a spatial gradient and
parameter $\epsilon$ appears in accordance to
equation (\ref{jtd}). 

We choose to work in the harmonic gauge \cite{fock-2} imposed on the metric tensor. It does not restrict physical applicability and interpretation of our subsequent results for the light ray deflection angle and the proper time delay which are directly observable and, hence, gauge-invariant quantities. In the linearized approximation the harmonic gauge condition is reduced to the Lorentz gauge, imposed on the potentials $\Phi$ and ${\bm A}$,
\begin{equation}
\label{3}
\frac{\epsilon}{c}\frac{\partial\Phi}{\partial t}+{\bm\nabla}\cdot{\bm A}=0\;.
\end{equation}
In this gauge the linearized Einstein equations can be written down as a system of the
gravimagnetic field equations \cite{m1}
\begin{eqnarray}
\label{4}
{\bm\nabla}\cdot{\bm E}&=&4\pi G\rho\;,\\
\label{5}
{\bm\nabla}\cdot{\bm B}&=&0\;,\\
\label{6}
{\bm\nabla}\times{\bm E}&=&-\frac{\epsilon}{c}\frac{\partial{\bm
B}}{\partial t}\;,\\
\label{7}
{\bm\nabla}\times{\bm B}&=&\frac{\epsilon}{c}\frac{\partial{\bm
E}}{\partial t}+\frac{4\pi G\epsilon}c{\bm j}\;,
\end{eqnarray}
where $\rho$ and ${\bm j}$ are the mass-density and mass-current of
the gravitating matter defined in terms of the energy-momentum tensor
$T_{\alpha\beta}$ of matter as $\rho=T^{00}/c^2$, and
$j^i=T^{0i}/c$ respectively \cite{LL,mtw}. 

Equations (\ref{1}), (\ref{3}), (\ref{4}) and (\ref{7}) are equivalent
to wave equations for the gravielectric and gravimagnetic potentials
\begin{eqnarray}
\label{a1}
\left(-\frac{1}{c^2_g}\frac{\partial^2}{\partial t^2}+\nabla^2
\right)\Phi(t,{\bm x})&=&-4\pi G\rho(t,{\bm x})\;,\\
\label{a2}
\left(-\frac{1}{c^2_g}\frac{\partial^2}{\partial t^2}+\nabla^2
\right){\bm A}(t,{\bm x})&=&-\frac{4\pi G}{c_{\rm g}}\,{\bm j}(t,{\bm x})\;.
\end{eqnarray}
These equations are of the hyperbolic type and describe the
propagation of the gravitational field with the speed
$c_{\rm g}=c/\epsilon$. Their physically relevant (causal) solution are 
the retarded potentials taken over volume of the body under consideration
\begin{eqnarray}
\label{a1r}
\Phi(t,{\bm x})&=&G\int_{V_{\rm body}}\frac{\rho(s,{\bm y})d^3 y}{|{\bm x}-{\bm y}|}\;,\\
\label{a2r}
{\bm A}(t,{\bm x})&=&\frac{G}{c_{\rm g}}\int_{V_{\rm body}}\frac{{\bm j}(s,{\bm y})d^3 y}{|{\bm x}-{\bm y}|}\;,
\end{eqnarray}
where
\begin{eqnarray}
\label{apo}
s=t-\frac{1}{c_{\rm g}}|{\bm x}-{\bm y}|\;,
\end{eqnarray}
is the retarded time due to the finite speed of propagation of gravity from the body to the observer (see Fig. \ref{fig_gwnc}). In the case of a point-like (or spherically-symmetric) massive body the point ${\bm y}\rightarrow{\bm z}$, where ${\bm z}$ is the coordinate of the body's center of mass. In this case the retarded time equation (\ref{apo}) is reduced to
\begin{eqnarray}
\label{apom}
s=t-\frac{1}{c_{\rm g}}|{\bm x}-{\bm z}(s)|\;,
\end{eqnarray}
where the body's center-of-mass coordinate ${\bm z}$ must be calculated at the retarded time $s$. This complicates solution of equation (\ref{apom}). Only in case of a uniform and rectilinear motion of the body, when ${\bm z}(s)$ is a linear function of time $s$, equation (\ref{apom}) can be solved exactly \cite{cqg}.   

Solutions (\ref{a1r}), (\ref{a2r}) are fully compatible with the matrix
$\Lambda^\alpha_\beta(\epsilon)$ of the Lorentz transformation
(aberration) of the gravitational field, which depends on the
fundamental speed $c_{\rm g}$ as well. Indeed, equations (\ref{a1}),
(\ref{a2}) are invariant with respect to the Lorentz transformation of
the space-time coordinates, the gravitational potentials and the
source variables using the matrix $\Lambda^\alpha_\beta(\epsilon)$
(see \S III and \S IV for more detail). Thus, measuring the
aberration of gravitational force would allow us to measure its relativistic properties such as the strength of gravimagnetic field \cite{kop-ijmpd}, 
the speed of gravity propagation \cite{k1,fk}, etc. In the case
of the instantaneous speed-of-gravity theory, where $c_{\rm g}=\infty$, equations (\ref{a1}), (\ref{a2}) are reduced to a
single Laplace-type equation for the scalar potential $\Phi$ and both
the gravimagnetic potential, ${\bm A}$, and the gravimagnetic
field, ${\bm B}$, vanish. The Lorentz invariance of the gravitational
field in such theory is totally broken and is not defined. 

The retarded potentials can be expanded in the post-Newtonian series. Formally, it is expansion in the Taylor series with respect to the parameter $\epsilon$ (that is, with respect to the speed of gravity $c_{\rm g}$). It yields
\begin{eqnarray}
\label{a1n}
\Phi(t,{\bm x})&=&G\int_{V_{\rm body}}\frac{\rho(t,{\bm y})d^3 y}{|{\bm x}-{\bm y}|}+O\left(\epsilon^2\right)\;,\\
\label{a2n}
{\bm A}(t,{\bm x})&=&\frac{G}{c_{\rm g}}\int_{V_{\rm body}}\frac{{\bm j}(t,{\bm y})d^3 y}{|{\bm x}-{\bm y}|}+O\left(\epsilon^3\right)\;.
\end{eqnarray}
Notice that the post-Newtonian expansion (\ref{a2n}) of the gravimagnetic potential ${\bm A}$ commences on the term being proportional to the speed of gravity $c_{\rm g}$. This is because the gravimagnetic field does not exist in the alternative theories of gravity where the gravity field propagates instantaneously, like the magnetic field would not exist in a hypothetical electromagnetic theory with the speed of light being equal to infinity \footnote{We keep the speed of light, c, constant}.

It is important to understand that the effects of the finite speed of gravity
reveal themselves not only through the retardation effects associated with second time derivatives in wave equations (\ref{a1}), (\ref{a2}), but also through the first time derivatives in equations (\ref{1})--(\ref{7}). It makes sense since the first time derivatives control the gravimagnetic and aberrational effects of the gravitational field in the near zone which are matched smoothly to the radiative-zone effects of emission and propagation of free gravitational waves generated by the isolated system. 

For a point-like massive body the mass current ${\bm j}=\rho{\bm v}$, and the integrals (\ref{a1n}), (\ref{a2n}) are reduced to
\begin{equation}
\label{8y}
\Phi=-\frac{GM}{r}+O\left(\epsilon^2\right)\;,\qquad\qquad{\bm
A}=\frac{\bm v}{c_{\rm g}}\Phi+O\left(\epsilon^3\right)\;,
\end{equation}
where $r=|{\bm r}|$, ${\bm r}={\bm x}-{\bm z}(t)$, and ${\bm z}(t)$ is
a coordinate of the body's center of mass in the global frame of reference. In section \ref{liab} we show that the potential ${\bm A}$ appears as a consequence of the Lorentz transformation of the gravitational field from the static frame of the body to the global (barycentric) frame. This establishes a correspondence between two ways of calculation of the gravimagnetic potential - directly from the field equations in the global frame, and by making use of the Lorentz transformation of the potential $\Phi$ from the static to a moving frame.   

\section{Lorentz Invariance and Aberration of Gravity}\label{liab}

The transformation between two reference frames moving with a constant
velocity with respect to each other, is described by the Lorentz
transformation.  The gravity field equations (\ref{a1}), (\ref{a2})
are invariant with respect to the Lorentz group with the matrix of
transformation $\Lambda^\alpha_\beta(\epsilon)$ having the ``standard" form
\cite{LL,mtw}
\begin{eqnarray}
\label{mat1}
\Lambda^0_{\;0}(\epsilon)&=&\gamma_\epsilon\equiv\left(1-\beta_\epsilon^2\right)^{-1/2}\;,\\
\label{mat2}
\Lambda^0_{\;i}(\epsilon)&=&\Lambda^i_{\;0}(\epsilon)=-\gamma_\epsilon\beta_\epsilon^i\;,\\
\label{mat3}
\Lambda^i_{\;j}(\epsilon)&=&\delta^{ij}+\left(\gamma_\epsilon-1\right)\frac{\beta_\epsilon^i\beta_\epsilon^j}{\beta_\epsilon^2}\;,
\end{eqnarray}
where the boost parameter $\beta_\epsilon^i=\epsilon v^i/c=v/c_{\rm g}$, and ${\bm
v}\equiv (v^i)$ is the relative velocity of observer with respect to the gravitating body. The matrix of the
inverse transformation
$\bar\Lambda^\alpha_\beta(\epsilon)=\Lambda^\alpha_\beta(-\epsilon)$. Moreover,
the transformation preserves the form of the Minkowski metric
\begin{equation}
\label{aaz}
\eta_{\alpha\beta}=\Lambda^\mu_{\;\alpha}(\epsilon)\Lambda^\nu_{\;\beta}(\epsilon)\eta_{\mu\nu}\;.
\end{equation}
The Maxwell equations 
are invariant with respect to the Lorentz group of electrodynamics with the matrix of
transformation $\lambda^\alpha_\beta$, given by \cite{LL,mtw}
\begin{eqnarray}
\label{mat1+}
\lambda^0_{\;0}&=&\gamma\equiv\left(1-\beta^2\right)^{-1/2}\;,\\
\label{mat2+}
\lambda^0_{\;i}&=&\lambda^i_{\;0}=-\gamma\beta^i\;,\\
\label{mat3+}
\lambda^i_{\;j}&=&\delta^{ij}+\left(\gamma-1\right)\frac{\beta^i\beta^j}{\beta^2}\;,
\end{eqnarray}
where the boost parameter $\beta^i= v^i/c$. The matrix of the
inverse transformation
$\bar\lambda^\mu_\nu(\beta)=\lambda^\mu_\nu(-\beta)$. Notice that $\lambda^\alpha_\beta=\Lambda^\alpha_\beta(\epsilon=1)$.

If $\epsilon\not=1$, the Lorentz transformation of the gravitational
field, given by the matrix $\Lambda^\alpha_\beta(\epsilon)$, is
different from that of the electromagnetic field equations, given by the matrix
$\lambda^\alpha_\beta$. Physically,
this means that gravitational interaction has the speed $c_{\rm g}$
that differs from the speed $c$ for electromagnetic field. In the
case of the instantaneous speed-of-gravity theory, when $\epsilon=0$,
the transformation (\ref{mat1})--(\ref{mat3}) degenerates to
$\Lambda^\alpha_{\beta}(\epsilon=0)=\delta^\alpha_{\;\beta}$ because
the space-time manifold is split in the Newtonian-like absolute time
and absolute space with a single gravitational potential $\Phi$
residing on it.

Lorentz transformation between two inertial frames, $x^\alpha$ and $x'^\alpha$, in gravity and electromagentic field equations are given by two different equations  
\begin{eqnarray}
\label{qzo1}
x'^\alpha&=&\Lambda^\alpha_\beta(\epsilon)x^\beta\;,\\
\label{qzo2}
x'^\alpha&=&\lambda^\alpha_\beta x^\beta\;.
\end{eqnarray}
Equation (\ref{qzo1}) transforms coordinates in the gravity field equations (\ref{a1}), (\ref{a2}) while equation (\ref{qzo2}) transforms coordinates in the Maxwell equations. It is important to observe that in the limiting case of a very slow velocity ${\bm v}$ the spatial part of the two boosts (\ref{qzo1}) and (\ref{qzo2}) is reduced to two Galilean transformations \cite{LL} that are not identical because $c_{\rm g}\not=c$. Specifically, equation (\ref{qzo1}) yields
\begin{equation}
\label{qzo4}
x'^i=x^i-\epsilon v^i t+O(\beta_\epsilon^2)\;,
\end{equation}
while the slow-velocity limit of equation (\ref{qzo2}) is reduced to
\begin{equation}
\label{qzo5}
x'^i=x^i- v^i t+O(\beta^2)\;.
\end{equation}
If one deals only with gravity field equations, the spatial coordinates $x^i$ can be rescaled,  $x^i\rightarrow \epsilon x^i$, so that the parameter $\epsilon=c/c_{\rm g}$ in equation (\ref{qzo4}) can be eliminated from linear terms and the difference between $c_{\rm g}$ and $c$ will reveal itself only in the post-Newtonian terms of order of $(v/c)^2$ and higher \footnote[2]{Equation (\ref{qzo4}) can be also presented in the Newtonian-like form by replacing time $t\rightarrow\tau=\epsilon t$. If $\epsilon\not=1$ the dynamic time $\tau=\epsilon t$ in gravity is different from atomic time $t$ in electrodynamics. Hence, measuring motion of photons in time-dependent gravitational field allows us to find out the presumable difference between the two times, $\tau$ and $t$, that is to measure $\epsilon=c/c_{\rm g}$ (see \cite{pla} for more detail).}. Gravitational light-ray deflection experiments observe interaction between light and gravity which depends on the spatial distance $|{\bm x}-{\bm z}|$ between coordinate ${\bm x}$ of the light particle and that ${\bm z}$ of a massive, light-ray deflecting body. Transformation from one to another reference frame changes their coordinates differently according to equations (\ref{qzo4}) and (\ref{qzo5}). Hence, the parameter $\epsilon$ can not be eliminated from the Galilean transformation of the relativite distance $|{\bm x}-{\bm z}|$ by the re-scaling of the spatial coordinates ${\bm z}$ of the body. Hence, it can be measured already in the linear post-Newtonian correction of order of $\epsilon v/c$ caused by translational motion of the body and appearing explicitly beyond the static part of the Shapiro time delay and/or standard Einstein's light-ray deflection equation \cite{carlip_comment,found-phys}.

Since the the current light deflection experiments are not sensitive
enough to measure the deflection of light caused by the body's spin,
we shall neglect the intrinsic gravimagnetic field associated with
the rotation of the body in the remainder of this paper.  This
approximation eliminates from equations (\ref{4})--(\ref{a2}) all
rotational currents.  (see \cite{cw} for more discussion of this
{\it intrinsic} rotation field).  The equations (\ref{4})--(\ref{a2}) solved in the
static frame of the light-ray deflecting body show that in this frame
the potentials, $\Phi'=(c^2/2)h'_{00}=-GM/r'$, ${\bm
A}'=-(c^2/4)h'_{0i}=0$, and the fields, ${\bm
E}'=-{\bm\nabla}\Phi'\not=0$, ${\bm B}'={\bm\nabla}\times{\bm A}'=0$,
where $r'=|{\bm x}'|$ is the radial coordinate of a field point in the
static frame, and $M$ is mass of the body. Transformation of the
static field potentials to the moving frame is obtained by applying
the Lorentz transformation to the metric tensor (gravitational field) perturbation
\begin{equation}
\label{arun}
h_{\alpha\beta}=\Lambda^\mu_{\;\alpha}(\epsilon)\Lambda^\nu_{\;\beta}(\epsilon)h'_{\mu\nu}\;,
\end{equation}
and is given more explicitly by equations
\begin{eqnarray}
\label{b1}
\Phi&=&\gamma^2_\epsilon\left[\left(1+\beta_\epsilon^2\right)\Phi'+4\left({\bm\beta}_\epsilon\cdot{\bm
A}'\right)\right]\;,\\
\label{b2}
{\bm A}&=&\gamma_\epsilon{\bm
A}'+\gamma^2_\epsilon\left[\Phi'+\frac{2\gamma_\epsilon+1}{\gamma_\epsilon+1}\left({\bm\beta}_\epsilon\cdot{\bm
A}'\right)\right]{\bm\beta}_\epsilon\;.
\end{eqnarray}
We emphasize that the gravitational potentials are transformed in accordance with the law of the Lorentz transformation of the gravitational field (\ref{arun}) so that it is the gravity null cone which enters these transformations. The light cone enters the law of the Lorentz transformation of the electromagnetic field (light). The presence of the gravity null-cone terms is traced by the parameter $\epsilon$ in all equations, and each physical effect directly associated with the finite speed of gravity can be unambiguously distinguished from that coupled to the light cone.  

Comparison with the PPN metric \cite{will} shows that our approach
formally yields the following values of the PPN parameters \cite{cqg}:
$\gamma=\beta=1$, $\alpha_1=8(\epsilon-1)$,
$\xi=\zeta_1=\zeta_2=\zeta_3=\zeta_4=\alpha_2=\alpha_3=0$. We
emphasize, however, that measurement of $\epsilon$ is not equivalent
to an independent measurement of $\alpha_1$ in the PPN formalism
framework \cite{will-r}, since the PPN formalism assumes that
$\epsilon=1$ while $\alpha_1\not=0$. This means that the two parameters have different physical origin (see next section) and coincide with each other accidentally due to the specific of the parameterizations adopted in \cite{will} and in the present paper.  Besides the measurement of
$\alpha_1$ is impossible without making an assumption
about the existence of a global preferred frame, usually associated
with the isotropy of the cosmic microwave background radiation (CMBR)
\cite{will-r}. Existence of such a ``preferred frame" in cosmology
should not be related to a possible violation of the Lorentz
invariance of gravitational field.  It simply confirms that our
universe is homogeneous and isotropic on large cosmological scales.
In contrast to the PPN formalism, the existence of the relativistic effects parametrized by $\epsilon$ do
not depend upon the ``preferred-frame" assumption. Evaluation of the
numerical value of $\epsilon$ itself is rendered through the
measurement of the {\it retardation of gravity} effect in the
gravitational time delay of light by moving bodies as shown in
\cite{cqg,k1,fk} and in equations (\ref{tdel})--(\ref{abt}) of this
paper.

The Lorentz transformation (\ref{b2}) generates the gravimagnetic
potential ${\bm A}$ and the gravimagnetic field ${\bm B}$ associated
with the translational motion of the (non-rotating) light-ray deflecting body.  Neglecting the
quadratic terms in equations (\ref{b1}), (\ref{b2}) one obtains in the
linear approximation that
\begin{equation}
\label{8}
\Phi=-\frac{GM}{r}+O\left(\epsilon^2\right)\;,\qquad\qquad{\bm
A}=\frac{1}{c_{\rm g}}\Phi{\bm v}+O\left(\epsilon^3\right)\;,
\end{equation}
where $r=|{\bm r}|$, ${\bm r}={\bm x}-{\bm z}(t)$, and ${\bm z}(t)$ is
a coordinate of the body's center of mass in the barycentric frame. We
emphasize that the Lorentz transformation (\ref{b1}), (\ref{b2}) of
the gravimagnetic potentials assumes that at each instant of time
the motion of the body is approximated by a straight line with
constant velocity ${\bm v}=d{\bm z}/dt$. We also bring the reader's attention that the equations (\ref{8}) are fully compatible with, and have the same physical content as the post-Newtonian solutions (\ref{a1n}), (\ref{a2n}) of the wave equations (\ref{a1}), (\ref{a2}). 

The Lorentz transformation of the static field ${\bm E}'$ of the body
generates the gravimagnetic field
\begin{equation}
\label{8a}
{\bm B}=\frac{1}{c_{\rm g}}\left({\bm v}\times{\bm E}\right)\;,
\end{equation}
where ${\bm E}=-{\bm\nabla}\Phi$. Equation (\ref{8a}) results from
definition (\ref{2}) and equation (\ref{8}).  The Lorentz transformation
(\ref{b1}), (\ref{b2}) describes change (aberration) in the arrangement
of the gravitational field lines of the body, measured in two different
frames. Equations (\ref{8}), (\ref{8a}) describe this {\it
gravimagnetic} effect of the {\it extrinsic} gravimagnetic field in
the linear approximation with respect to the parameter $\epsilon=c/c_{\rm g}$.

To summarize this section, in classical electrodynamics a uniformly
moving charge generates magnetic field. This is because the Lorentz
transformation generates electric current which produces the magnetic
field. The resulting magnetic field is real, since it couples to electric field in a gauge-invariant way, and can be measured. Its
observation confirms that electromagnetic field is Lorentz-invariant
and its speed of propagation is $c$ \cite{jackson}.  Similarly, the
gravimagnetic potential (\ref{8}) and the gravimagnetic field
(\ref{8a}) are solutions of the Einstein equations and lead to gauge-invariant observable effects which can be
measured in gravitational experiments. It provides a test of the
Lorentz invariance of gravitational field and measurement of the
fundamental speed of gravity $c_{\rm g}=c/\epsilon$ which controls the causal property of gravity (see section \ref{cslt}) and the gravimagnetic effects.

\section{Lorentz Transformation of Gravity and Light in the Shapiro Time Delay} 
\subsection{Equation of Light Propagation and Its Lorentz Transformation}

In the general theory of relativity, the equations of motion for light
propagating in vacuum are null geodesics, and
the general relativistic equations of light propagation parameterized by
coordinate time $t$ are \cite{cqg}
\begin{equation}
\label{lg}
\frac{d^2x^i}{ dt^2}=c^2k^\mu k^\nu\left(k^i
\Gamma^0_{\mu\nu}-\Gamma^i_{\mu\nu}\right)\;,
\end{equation}
where $k^\mu=(1,{\bm k})$ is a null four-vector ($\eta_{\mu\nu}k^\mu
k^\nu=0$) of a photon, and ${\bm k}=(k^i)$ is the unit Euclidean vector that is
tangent to the unperturbed photon trajectory. The null vector
$k^\mu$ is transformed in accordance to the Lorentz group of
the transformation of electromagnetic field
\begin{equation}
\label{pox}
K^\mu=\bar\lambda^\mu_{\;\nu}k^\nu\;,
\end{equation}
where $K^\mu$ and $k^\mu$ are components of the null vector of photon
in the static and moving reference frames, respectively, and
$\bar\lambda^\mu_{\;\nu}$ is the matrix inverse with respect to
$\lambda^\mu_{\;\nu}$.  The Levi-Civita connection describes the force
of gravitational attraction and is transformed in accordance with the
Lorentz group of transformation of gravitational field, that is
\begin{equation}
\label{xru}
\Gamma^\alpha_{\beta\gamma}=\Lambda^\alpha_{\;\sigma}(-\epsilon)\Lambda^\mu_{\;\beta}(\epsilon)\Lambda^\nu_{\;\gamma}(\epsilon)\Gamma'^\sigma_{\mu\nu}\;.
\end{equation}
If the parameter $\epsilon\not=1$, the Lorentz group of
electromagnetic field does not coincide with that of gravitational
field. Hence, the equation (\ref{lg}) must be modified in order to
keep it form-invariant. It is beyond the scope of this paper to derive
this generalized equation of light propagation in a particular theory
of gravity with broken Lorentz invariance of gravitational field. This problem has been tackled in our papers \cite{found-phys,kwtn}. The goal of the present paper is to test whether the Lorentz group of
transformation of gravitational field 
coincides with that of electromagnetic field. To achieve this goal we
assume that equation (\ref{lg}) is valid in the observer frame with respect to which a
light-ray deflecting body is moving. Then, we take electromagnetic and gravitational field variables in the static frame of the body and apply the Lorentz transformation with $\epsilon=1$ for light and $\epsilon\not=1$ for gravity,
thus,
obtaining their transformed values which are substituted to
equation (\ref{lg}). The obtained equation is parametrized by
parameter $\epsilon$, and measuring
$\epsilon$ in gravitational time-delay experiments allows us to measure the speed of gravity $c_{\rm g}$ with respect to the speed of light.

Final notice relates to the nature of the light-ray geodesic equation (\ref{lg}). Some researchers (see \cite{will-r,will-web} and references therein) state that because this equation describes propagation of light it can not contain any information about the speed with which gravity propagates. Such statement is valid only in case when the light-ray deflecting body is at rest both with respect to observer and to the source of light. However, the right side of equation (\ref{lg}) is the gravity force acting on a light particle, and in case when the body is moving, the force of gravity does not propagate instantaneously from the body to the light particle. This propagation of gravity is in the form of the "acceleration-dependent" gravitational field when the light particle moves in a far (radiative) zone of the body where transverse-traceless gravitational waves dominate. In the near (non-radiative) zone of the body's gravitational field the gravitational interaction propagates to the light particle in the form of the "velocity-dependent" gravitational field. The acceleration-dependent and the velocity-dependent components of gravitational fields always present together in the most general case of accelerated worldline of the light-ray deflecting body. In case of a uniformly moving body, the acceleration-dependent part of gravitational field of the body is absent and transverse-traceless gravitational waves are not emitted. The radiative zone is absent and the light particle always moves in the near zone of the body described by its velocity-dependent component. But it does not mean that the inetraction of the light particle with the velocity-dependent component is instantaneous. The gravitational field still obeys the causal property and the light particle can "feel" the gravitational field only after it crosses the gravity null cone as shown in Fig. \ref{buh} which is based on the result of the integration of the light-ray propagation equation (\ref{lg}) as shown in section \ref{timed}. 

Some researchers also states erroneously that the principle of equivalence tells us nothing about the speed of gravity \cite{will-r,will-web}. This is because the principle of equivalence  eliminates gravitational force acting on a test particle at each point of its world line by making a local coordinate transformation to a free-falling coordinate system. This statement is true, but it is not applicable to the light-ray deflection experiments which are essentially non-local and, as such, test the global properties of the gravitational field which can not be measured in local experiments. Gravitational field can not be eliminated along the entire light-ray trajectory by a single coordinate transformation. Therefore, when the body is moving, a light particle propagates through non-zero, time-dependent gravitational field which can not interact with the particle instantaneously if the speed of gravity is finite. The light particle (photon) propagates through a sequence of the gravity null cones depicted in Fig. \ref{buh} which describes the interaction of the light particle with the gravitational field of the body in the background flat space-time. Gravitational light-ray deflection angle and time delay are integrals taken along the unperturbed light null cone but the gravity null cone eneters this calculation as well through the time derivatives in the Christoffel symbols normalized to the speed of gravity $c_{\rm g}$ (see next section). Our calculations reveal, that numerical values of the time delay and the deflection angle of the light particle, after it passes the light-ray deflecting body, are determined at each point of the light-ray trajectory (for instance, points 3, 4, 5 in Fig. \ref{buh}) by the retarded positions of the body (positions C, D, E in Fig. \ref{buh}) connected to the light particle by the gravity null cone. The retarded position of the body must not be confused with the point of the closest approach of the light particle to the body. The point of the light's closest approach lies on the light null cone and can be spatially close to the retarded position of the body at the time of observation. However, the equation defining the point of the closest approach does not follow from the result of the integration of the light ray in the gravitational field of a moving body, and appears nowhere in the mathematical description of the observable quantities like the time delay and the total deflection angle of light.  

This mathematical prediction of the retarded position of the body in the expressions for the time delay and the deflection angle follows from the causal nature of the gravity field equations, and it leads to the inevitable conclusion that at the time of observation, $t$, the photon received by observer (point 5 in Fig. \ref{buh}) is deflected and delayed by the body's gravitational field from the retarded position (point E in Fig. \ref{buh}) of the body, $z(s)$, taken on the "last" gravity null cone  (see equations (\ref{tdel}) and (\ref{abt} below). If the speed of gravity $c_{\rm g}$ is infinite, the body deflects (and delays) light by its gravitational field taken at the time $t$, when it reaches observer, which means that the gravity propagates information about position of the body instantaneously, and does not obey the causality principle. This consideration reveals that the gravity null cone (determined by the speed of gravity) is an essential part of the light-ray deflecting experiments conducted in the field of a moving massive body (see section \ref{cslt} and Fig. \ref{fig_lgwnc} for further details) which space-time structure can be measured in the ultra-precise observation of the deflection of light coming to observer from a star/quasar. This study refutes the arguments presented in \cite{will-r,will-web} that are based on misconceptions in understanding of the interaction of electromagnetic and gravitational fields.

\subsection{Lorentz Transformation of the Levi-Civita Connection}

The Levi-Civita connection $\Gamma'^\alpha_{\mu\nu}$ in a static frame
is given by the equations
\begin{eqnarray}
\label{ss2}
\Gamma'^0_{00}&=&\Gamma'^0_{ij}=\Gamma'^i_{0j}=0\;,\\
\label{ss3}
\Gamma'^0_{0i}&=&\Gamma'^i_{00}=-\frac{\partial\Phi'}{\partial x'^i}\;,\\
\label{ss7}
\Gamma'^i_{jp}&=&-\delta_{jp}\frac{\partial\Phi'}{\partial
x'^i}+\delta_{ip}\frac{\partial\Phi'}{\partial x'^j}+
\delta_{ij}\frac{\partial\Phi'}{\partial x'^p}\;,
\end{eqnarray}
where $\Phi'=(c^2/2)h'_{00}=-GM/c^2r'$, and $r'=|{\bm x}'|$. As
follows from equations (\ref{ss2})--(\ref{ss7}), the Levi-Civita
connection is associated in the static frame with the force of
gravitational attraction \cite{LL,mtw}: it is {\it not} associated
with electromagnetism. Therefore, its transformation in equation
(\ref{xru}) from the static to a uniformly moving frame must be done
with the matrix $\Lambda^\alpha_{\;\beta}(\epsilon)$ pertained to the
Lorentz transformation of the gravitational field.  Substituting the
matrix of the Lorentz transformation (\ref{mat1})--(\ref{mat3}) to
equation (\ref{xru}) and making use of equations
(\ref{ss2})--(\ref{ss7}) along with equations (\ref{b1}) and
(\ref{b2}), yield the components of the Levi-Civita connection in the
moving frame
\begin{eqnarray}
\label{bb2}
\Gamma^0_{00}&=&-\frac{1}{c_{\rm g}}\frac{\partial\Phi}{\partial t}\;,\\
\label{bb3}
\Gamma^0_{0i}&=&-\frac{\partial\Phi}{\partial x^i}\;,\\
\label{bb4}
\Gamma^0_{ij}&=&+2\left(\frac{\partial A^j}{\partial x^i}+
\frac{\partial A^i}{\partial
x^j}\right)+\frac{1}{c_{\rm g}}\frac{\partial\Phi}{\partial
t}\delta_{ij}\;,\\
\label{bb5}
\Gamma^i_{00}&=&- \frac{\partial\Phi}{\partial
x^i}-\frac{4}{c_{\rm g}}\frac{\partial A^i}{\partial t}\;,\\
\label{bb6}
\Gamma^i_{0j}&=&-2\left( \frac{\partial A^i}{\partial
x^j}-\frac{\partial A^j}{\partial
x^i}\right)+\frac{1}{c_{\rm g}}\frac{\partial\Phi}{\partial
t}\delta_{ij}\;,\\
\label{bb7}
\Gamma^i_{jp}&=&-\delta_{jp}\frac{\partial\Phi}{\partial
x^i}+\delta_{ip}\frac{\partial\Phi}{\partial x^j}+
\delta_{ij}\frac{\partial\Phi}{\partial x^p}\;,
\end{eqnarray}
where the parameter $\epsilon=c/c_{\rm g}$ explicitly appears in front of the time
derivatives in equations (\ref{bb2}), (\ref{bb4})--(\ref{bb6}) as a
direct consequence of the Lorentz transformation law (\ref{xru}). Its
appearance complies with the phenomenological parametrization rule in
equation (\ref{jtd}), tracking the presence of the fundamental speed
of gravity in various gravitational equations.

It is important to compare equations (\ref{bb2})--(\ref{bb7}) with the
definition of the Levi-Civita connection adopted in the PPN formalism
\cite{will} which postulates that the transformation of the connection
obeys the ``electromagnetic" rule
\begin{equation}
\label{xppn}
\Gamma^\alpha_{\beta\gamma}=\bar\lambda^\alpha_{\;\sigma}\lambda^\mu_{\;\beta}\lambda^\nu_{\;\gamma}\Gamma'^\sigma_{\mu\nu}\;,
\end{equation}
where $\lambda^\alpha_{\;\sigma}$ is the matrix of the Lorentz
transformation of electromagnetic field. Straightforward calculations
reveal that
\begin{eqnarray}
\label{pp2}
\Gamma^0_{00}&=&-\frac{1}c\frac{\partial\Phi}{\partial t}\;,\\
\label{pp3}
\Gamma^0_{0i}&=&-\frac{\partial\Phi}{\partial x^i}\;,\\
\label{pp4}
\Gamma^0_{ij}&=&+2\left(\frac{\partial A^j}{\partial x^i}+
\frac{\partial A^i}{\partial
x^j}\right)+\frac{1}c\frac{\partial\Phi}{\partial t}\delta_{ij}\;,\\
\label{pp5}
\Gamma^i_{00}&=&-
\frac{\partial\Phi}{\partial x^i}-\frac{4}c\frac{\partial A^i}{\partial t}\;,\\
\label{pp6}
\Gamma^i_{0j}&=&-2\left( \frac{\partial A^i}{\partial
x^j}-\frac{\partial A^j}{\partial
x^i}\right)+\frac{1}c\frac{\partial\Phi}{\partial t}\delta_{ij}\;,\\
\label{pp7}
\Gamma^i_{jp}&=&-\delta_{jp}\frac{\partial\Phi}{\partial
x^i}+\delta_{ip}\frac{\partial\Phi}{\partial x^j}+
\delta_{ij}\frac{\partial\Phi}{\partial x^p}\;,
\end{eqnarray}
which should be compared with equations (\ref{bb2})--(\ref{bb7}).  One
can see that the PPN-defined Levi-Civita connection
(\ref{pp2})--(\ref{pp7}) assumes explicitly that the parameter
$\epsilon=1$, which excludes it from the set of the parameters that
probes the relativistic transformation properties of the gravitational force (the Levi-Civita connection).
 
Obviously, our $\epsilon$-parametrization of the Einstein equations
\cite{cqg} will make {\it all} time derivatives of the gravitational
potentials vanish if the speed of gravity $c_{\rm g}\rightarrow\infty$.  This is intuitively pleasing since the
limit $c_{\rm g}\rightarrow\infty$ corresponds to the case of the infinite
speed-of-gravity theory where the gravitational interaction is
instantaneous, and both the gravity field equations and equations of
motion of test particles, which depend on the Levi-Civita connection,
should not contain any time derivatives of the gravitational field.

In contrast, the PPN formalism \cite{will} demands that the equations
of motion of test particles depend on the connection
(\ref{pp2})--(\ref{pp7}) that contains non-vanishing first time
derivatives of the gravitational field even in the limit
$c_{\rm g}\rightarrow\infty$ because equations (\ref{pp2})--(\ref{pp7}) do
not depend on the fundamental speed of gravity $c_{\rm g}$ explicitly. In
other words, the PPN formalism orders that the gravitational force
propagates with the speed of light and in this framework, the speed of gravity cannot be tested.  We, thus, consider the PPN equations
(\ref{pp2})--(\ref{pp7}) essentially incomplete because
the degree of freedom corresponding to the propagation of
gravitational force that may differ from the speed of light is
missing.  

\subsection{Gravitational Time Delay and Deflection of Light}\label{timed}  

The double integration of equation (\ref{lg}) along the unperturbed light ray
 yields the time of propagation of light from the point $x_0^i$ to the point, $x^i=x^i(t)$,  on the trajectory of the light ray 
\begin{equation}
\label{qer}
t-t_0=\frac{1}{ c}|{\bm x}-{\bm x}_0|+\Delta(\epsilon)\;,
\end{equation}
where the relativistic time delay \cite{cqg}
\begin{equation}
\label{aa}
\Delta(\epsilon)=\frac{c}{ 2}\;k^\mu k^\nu
k_\alpha\int_{t_0}^{t}d\tau\int_{-\infty}^\tau d\sigma
\Bigl[\Gamma^\alpha_{\mu\nu}(\sigma,{\bm x})\Bigr]_{{\bm x}={\bm
x}_N(\sigma)}
\end{equation}
is a function of the coordinates of emission, $x_0^i$, and the point, $x^i$, where the light particle is located at the time $t$.
Integration in equation (\ref{aa}) is along a straight line of unperturbed
propagation of the photon
\begin{equation}
\label{und}
{\bm x}_N(t)={\bm x}_0+c{\bm k}(t-t_0)\;,
\end{equation}
where $t_0$ is time of emission, ${\bm x _0}$ is the coordinate of the
source of light at time $t_0$. By substituting equations
(\ref{bb2})--(\ref{bb7}) into the time delay equation (\ref{aa}) and
making use of the gauge condition in equation (\ref{3}) in order to
replace the time derivative of the potential $\Phi$ to the divergence
of the potential ${\bm A}$, we obtain the following form
\begin{equation}
\label{ty}\fl
\Delta(\epsilon)=2\int_{t_0}^{t}\Phi(\tau,{\bm x}_N(\tau)) d\tau
-2\left(1-\frac{1}{\epsilon}\right)\int_{t_0}^{t}d\tau\int_{-\infty}^{\tau}\Bigl[
{\bm\nabla}\cdot{\bm A}(\sigma,{\bm x})\Bigr]_{{\bm x}={\bm
x}_N(\sigma)}d\sigma+O\left(\delta^2\right)\,,
\end{equation}
where $\delta\equiv\epsilon-1$ and all terms of order $\delta^2$ or
higher are omitted.  The potentials $\Phi$ and ${\bm A}$ are given by
equation (\ref{8}) and are taken on the unperturbed light-ray
trajectory (\ref{und}); that is, ${\bm x}={\bm x}_N(t)$. We emphasize
that the standard PPN formalism, with $\epsilon=1$, ignores the possible presence of the second
term in the right side of equation (\ref{ty}) which comes from
the explicit dependence of $\Gamma^\alpha_{\mu\nu}$ on $c_{\rm g}$ in the
right side of equations (\ref{bb2})--(\ref{bb7}). For this reason the PPN formalism
is not sensitive to violation of the Lorentz-invariance of the
gravitational force associated with the transformation of the
Levi-Civita connection \cite{cqg}, and is helpless in physical interpretation of the gravitational light-ray deflection experiments.

When light propagates through a gravitational field, it is not only
delayed but deflected as well. Let us denote $\alpha^i$ the angular
coordinates of the deflection vector referred to the plane of the sky
that is orthogonal to the unit vector $k^i$ that defines propagation
of the unperturbed light ray.  In general relativity, the deflection
vector $\alpha^i$ is defined as a spatial derivative of the
relativistic perturbation of the electromagnetic phase \cite{LL,mtw}
that is proportional to the time delay (\ref{ty}), plus
the observer-dependent relativistic correction taking into account
non-integral (local) interference of the light-ray particle with the
gravitational field perturbations \footnote[3]{ Exact Lorentz-invariant equation for the total light-ray deflection angle $\alpha^i$ in the framework of general relativity for the case $\epsilon\equiv 1$ has been derived in our paper \cite{ks} and is given by equation (67) in there.}. We use this
definition in our parametrized approach, and define the parametrized value of the deflection angle as
\begin{equation}
\label{da}
\alpha^i(\epsilon)=-cP^{ij}\left[\frac{\partial\Delta(\epsilon)}{\partial
x^i}-k_\alpha h^{i\alpha}\right]\;,
\end{equation}
where $P^{ij}=\delta^{ij}-k^i k^j$ is operator of projection on the
plane of the sky orthogonal to vector $k^i$.

\subsection{Characteristic Time Interval and Spatial Domain of the Gravitational Deflection of Light}

In the evaluation of equation (\ref{ty}), the duration of the time
integration interval is important for better understanding of the
gravitational physics of the light-ray deflection experiments.  The
region of space and time interval during which the interaction of light
particle with gravitating body is relevant can be estimated as
follows: A light particle is deflected (scattered) by a long-range
gravitational force whose potential $\Phi$ decreases basically as
$1/r$, where $r$ is the distance of the light particle from the
light-ray deflecting body. The approximate expression for the
deflection angle $\alpha$ of a light particle emitted at infinity is
obtained from equation (\ref{da}), if one neglects motion of the light
ray-deflecting body. For an observer located behind the body with
respect to the source of light, the result is \cite{LL,mtw}
\begin{equation}
\label{deh}    
\alpha(r)=\frac{4GM}{c^2d}\frac{r}{\sqrt{r^2-d^2}}\;,
\end{equation}
where $d$ is the impact parameter of the light ray with respect to the
body that is fixed in space at some instant of time \footnote[4]{ More precisely, the impact parameter of the light ray is the spatial distance of the closest approach of the ray to the light-ray deflecting body. In general, it depends on the coordinate frame used for calculations.}.  Let us assume that the astrometric accuracy for the
angular measurements is $\Delta\alpha$.  Then, the region of radius
$r_s$ from the body where the gravitational interaction of light with
the body is significant is determined by solving equation
\begin{equation}
\label{wq1}    
\alpha_\infty-\alpha(r_s)=\Delta\alpha\;,
\end{equation}
where $\alpha_\infty=4GM/c^2d$.  We assume that $d<r_s$ so that we can
solve equation (\ref{wq1}) by expanding it in a Taylor series with
respect to the parameter $d/r_s$, which gives
\begin{equation}
\label{wq2}
r_s\simeq d\left(\frac{\alpha_\infty}{2\Delta\alpha}\right)^{1/2}\;.
\end{equation}
The time that it takes light to cross this region is
\begin{equation}
\label{wq3}
t_s=\frac{2r_s}{c}\simeq
\frac{d}{c}\left(\frac{2\alpha_\infty}{\Delta\alpha}\right)^{1/2}\;,
\end{equation}
which gives us a characteristic time of the gravitational interaction
of light with the body under consideration, assuming an angular
tolerance of $\Delta\alpha$.
 
Current VLBI technology can measure changes in the relative angles
between sources in the sky with an accuracy $\sim$10 $\mu$arcsec
\cite{fk}, and in the next decade the accuracy may approach $\sim$ 1
$\mu$arcsec $=5\times 10^{-12}$ rad \cite{sim,gaia,ska,freid}. As an example,
consider the Sun as a light-ray deflecting body with the light-ray
impact parameter $d=R_\odot=7\times 10^{10}$ cm -- the radius of the
Sun. Using a tolerance of $\Delta\alpha= 10$ $\mu$arcsec, we obtain
from equation (\ref{wq2}) that $r_s\simeq 300 R_\odot=2.1\times
10^{13}$ cm. This means that a moving particle of light continuously
interacts with the gravitational field of the Sun in a region around
the Sun with the size which is slightly above one astronomical unit (1
AU = $1.5\times 10^{13}$ cm). The light travel time across this region
is $t_s\sim 23$ minutes which gives the characteristic time of
interaction of light particle with the gravitational field of the Sun
for an assumed astrometric accuracy. Similar calculation for the light
particle moving in the gravitational field of Jupiter and grazing its
limb, shows that the characteristic time of light scattering is about
$t_s\simeq 14$ seconds, and the region of the gravitational interaction
of light with Jupiter is about $r_s\simeq 30$ jovian radii.

These estimates elucidate that the characteristic time of interaction
of electromagnetic signal and gravity in the light-ray deflection
experiments is sufficiently large so that this interaction can not be
considered as an instantaneous scattering by a point-like mass to
produce a total deflection angle $\alpha_\infty$. This ``instantaneous-scattering" point of view
about the gravitational deflection of light is commonly used in the
gravitational lensing theory to map images of the sources of light
from the source plane to observer \cite{peter}. However, it may be a
misleading concept for analysis of the light-ray deflection
experiments in the solar system.

\section {Summary of the Physical Interpretations of the Deflection Experiments in the Field of a Moving Gravitational Body}

\subsection{Significance of the Velocity-Dependent Corrections Caused by Motion of the Body} 

Relativistic corrections caused by the motion of the light-ray
deflecting body are directly associated with transformation properties
of the gravitational field regulated by the fundamental speed of
gravity $c_{\rm g}$ which defines the null cone of gravity propagation.  For example, consider the relativistic deflection of
light caused by the Sun and viewed from the geocentric frame. The Sun
moves in this frame with velocity $v_\odot\simeq 30$ km/s and
acceleration $a_\odot\simeq 0.6$ cm/s$^2$. Let us also assume that
position of the Sun ${\bm z}(t)$ is fixed at some instant of time
$t_A$, ${\bm z}(t_A)$, and that the total angle of the deflection of
light is given by equation for a static gravitating body:
$\alpha=4GM_\odot/c^2d$, where $d$ is the impact parameter of the
light ray with respect to the body referred to the time $t_A$.

A light ray grazing the solar limb ($d\simeq R_\odot$) interacts with
the Sun during time $t_s\le 23$ minutes for an assumed angular
accuracy of $10~\mu$arcsec.  Hence, if one ignores the geocentric
velocity of the Sun it will produce an error $\delta d=v_\odot t_s$ in
the value of the impact parameter, which leads to error
$\delta\alpha_v$ in calculation of the angle of the gravitational
light deflection of $\delta\alpha_v\sim\alpha_\odot(v_\odot
t_s/R_\odot)\simeq 105$ mas, where $\alpha_\odot=1.75''$. The
geocentric acceleration of the Sun, $a_\odot=\dot v_\odot$, introduces
an error $\delta d=a_\odot t^2_s/2$ to the impact parameter and an
error $\delta\alpha_a\sim\alpha_\odot(a_\odot t^2_s/2R_\odot)\simeq
15$ $\mu$arcsec to the deflection angle \footnote[5]{Proper physical treatment of the acceleration-dependent
relativistic effects in the propagation of light ray requires more
extended mathematical calculations (see, for example, \cite{ks}).}.  These estimates
clearly show that, if precision of angular measurements is approaching
10 $\mu$arcsec, the relativistic deflection of light and the Shapiro
time delay in the geocentric frame are affected by the geocentric
velocity and acceleration of the Sun and, hence, must be taken into
account for adequate physical interpretation of the results of the
solar gravity-deflection experiments \footnote[6]{The magnitude of the acceleration-dependent terms in the light-ray deflection experiments shown in \cite{will-web} is at the thousandths of a picosecond level ($< 10^{-4}$ $\mu$arcsec). We emphasize that this estimate relates to terms of higher order than those taken into account in the present paper.}.

Because both light and the gravitating body are moving as the light
propagates, the instant of time $t_A$ associated with the impact
parameter $d$ (when the deflection is greatest) requires precise
evaluation of appropriate time delay and deflection integrals.  Some
claims state \cite{hellings} 
that this instant of time is the time of the closest approach of
photon to the light-ray deflecting body. However, these claims were
based rather on an intuitive guess than on precise mathematical
calculation of $t_A$.  The appropriate calculation shown below in equations
(\ref{pok})--(\ref{abt}) for the time delay in the field of a moving gravitating
body (see \cite{cqg,k1,found-phys}) shows that it deflects light from the
retarded position ${\bm z}(t_A)={\bm z}(s)$ on its world line where
the retarded time $s$ is defined by equation (\ref{a5}) for null
characteristics of the gravitational field produced by a moving
body. This result is in a full agreement with general relativistic
prediction that the causality principle for gravity is valid and that
the gravity propagates and interacts with other particles on the null cone.

\subsection{Time Delay and Deflection of Light by a Moving Body}

Integration of equation (\ref{ty}) for the time delay can be performed
with sufficient accuracy under the assumption of uniform and
rectilinear motion of the light-ray deflecting body with constant
velocity ${\bm v}$, that is
\begin{equation}\label{pok}
{\bm z}(t)={\bm z}(t_{\rm A})+{\bm v}(t-t_{\rm A})+O(t-t_{\rm A})^2\;,
\end{equation}
where ${\bm z}(t_{\rm A})$ is the position of the body at time $t_{\rm
A}$, with quadratic and acceleration-dependent terms neglected. Light propagates through time-dependent gravitational field as shown in Fig. \ref{buh} and the time delay is given by an integral along the light-ray trajectory. The
evaluation of the integral in the time delay is tedious but
straightforward, and exhaustive description of this calculation can be
found in sections 3 and 4 of our paper \cite{cqg}, so they are not
repeated here. The result is given by \cite{cqg,pla}
\begin{eqnarray}
\label{tdel}
\Delta(\epsilon)&=&-(1+\gamma)\left(1-\frac{1}{c_{\rm g}}{\bm k}\cdot{\bm
v}\right)\frac{GM}{c^3}\ln\left(r-{\bm k}\cdot{\bm
r}\right)+O\left(\delta^2\right)\;,
\end{eqnarray}
where $\delta\equiv\epsilon-1$, ${\bm r}={\bm x}-{\bm z}(s)$, ${\bm
v}=d{\bm z}(s)/ds$ is velocity of the body, and the coordinates and
velocity of the light-ray deflecting body in equation (\ref{tdel}) are
taken at the retarded time $t_A=s$ where
\begin{eqnarray}
\label{a5}
s&=&t-\frac{1}{c_{\rm g}}|{\bm x}-{\bm z}(s)|\;.
\end{eqnarray}
Equations (\ref{tdel}), (\ref{a5}) were used to draw Fig. {\ref{buh}. It shows that the relativistic time delay of photon at each consequtive point on its trajectory is determined by the position ${\bm x}$ of the photon at time $t$, and the retarded position of the body ${\bm z}(s)$ on the gravity null cone with vertex of the gravity cone on the world line of the body taken at the retarded time $s$ in accordance with equation (\ref{a5}). Notice that neither the point nor the time of the closest approach of the photon to the body appear in the expression (\ref{tdel}) for the time delay.   

From now on, we shall treat the point ${\bm x}$ on the light-ray trajectory as taken at the point of observation, and time $t$ as the time of observation (point 5 in Fig. \ref{buh}). This is because we can not make any measurement of photon before it arrives to observer.
It is remarkable that the retardation time $s$ of gravity is determined only by the "last" gravity null cone connecting the body and the observer at the time of observation of the photon $t$. For this reason, we do not show in Figs. (\ref{fig_lgwnc}) and (\ref{fig_lgwnc}) the entire sequence of the gravity null cones interacting with light like we did in Fig. \ref{buh}. 
The appearance of the speed of gravity $c_{\rm g}$ in equation (\ref{a5}) reflects the fact that the gravity force generated by the moving body does not interact with light instantaneously but with a finite speed because of its causal nature. Had the gravity propagated instantaneously with $c_{\rm g}=\infty$, the time delay equation (\ref{tdel}) would depend on the position of the light-ray deflecting body taken at the time of observation $t$. Graphical presentation of the gravitational physics lying behind the time delay equation (\ref{tdel}) is shown in Fig. \ref{buh} and \ref{fig_lgwnc} (see also graphics in \cite{found-phys}). 

The light-ray deflection vector in the plane of the sky is calculated
from equation (\ref{da}) and is given by
\begin{eqnarray}
\label{abt}\fl
\alpha^i(\epsilon)&=&(1+\gamma)\frac{GM}{c^2}\frac{\left(1-\displaystyle{\frac{1}{c_{\rm g}}}{\bm
k}\cdot{\bm v}\right)^2}{r-\displaystyle{\frac{1}{c_{\rm g}}}{\bm v}\cdot{\bm
r}}\frac{\xi^i}{r-{\bm k}\cdot{\bm r}}-(1+\gamma)\frac{2GM}{c^2r}\frac{v^i_T}{c_{\rm g}}
+O\left(\delta^2\right)\;,
\end{eqnarray}
where $\xi^i=P^i_{\;j}r^j$ is the impact distance of the light ray to
the moving body taken at the retarded instant of time, and
$v^i_T=P^i_{\;j}v^j$ is the transversal velocity of the body lying in
the plane of the sky orthogonal to vector $k^i$.  For generality, we
have added the PPN parameter $\gamma$ in equations (\ref{tdel}) and
(\ref{a5}) to include possible violations of general relativity,
caused by presence of hypothetical scalar fields remained
after epoch of the cosmological inflation \cite{DN}.  The addition of
$\gamma$ does not interfere with the Lorentz transformations
\cite{ks,kv} or contradicts any of the previous formulations.

We emphasize that the velocity-dependent effects (i.e. $v/c_{\rm g}$) appear both {\it
explicitly} in the time delay and light-ray deflection equations
(\ref{tdel}), (\ref{abt}) and {\it implicitly} through the retarded
time equation (\ref{a5}) that displaces the position of the body on
its world-line backward in time with respect to the instant of
observation $t$. This occurs because equation (\ref{a5}) is a null
characteristic of the gravitational field equations (\ref{a1}),
(\ref{a2}) describing the relativistic effect of the propagation of
gravity \cite{k1}. The appearance of $c_{\rm g}$ in the time delay equation (\ref{tdel}) is a
natural consequence of the fact that the Levi-Civita connection
(\ref{bb2})--(\ref{bb7}) contains time derivatives which are
normalized to the speed of gravity $c_{\rm g}=c/\epsilon$ in accordance with
the Lorentz-invariant properties of the gravitational field that
controls the propagation of gravity force both in the near and radiative zones in the form of free gravitational waves 
\cite{found-phys,bs}. In other words, the Lorentz invariance of gravity and its
finite speed of propagation are physically tightly connected and should
not be conceptually separated from each other.

The physical interpretations associated with the light deflections
experiments in the gravitational field of moving bodies are also
discussed elsewhere (see, for instance,
\cite{cqg,pla,fk,found-phys}).  However, in the remainder of this
section we summarize them for completeness and for the reader's convenience.

\subsection{Relativistic Delay Measurement of the Gravimagnetic Field}

Equation (\ref{tdel}) describes the relativistic time delay of light
(radio waves) caused by a massive body moving with velocity ${\bm v}$
with respect to the rest frame of observer located at the point ${\bm
x}$. Gravity is a long-ranged field and the time delay given by
equation (\ref{ty}) is effectively an integral of the
gravitational force exerted on the photon along its entire
trajectory. However, the strongest impact of the time-dependent
gravitational field on the photon emitted at time $t_0$ is when the
massive body is located in its retarded position ${\bm z}(s)$ taken at
the retarded time $s$, determined by equation (\ref{a5}). If general
relativity is correct, then, $c_{\rm g}=c$ and only the gravielectric
potential $\Phi$ of the body is essential for calculation of the
time delay (\ref{tdel}). However, if $c_{\rm g}\not=c$ 
(hence, general relativity is not correct), then the gravimagnetic
potential ${\bm A}$ adds a contribution to the delay, as shown by
equation (\ref{ty}). 
The second term (the double integral) in the right side of this equation
is produced by the gravimagnetic field due to the orbital
motion of the light-ray deflecting body. If general relativity is correct, the contribution of
the double integral in equation (\ref{ty}) to the time delay is
identically zero, and the magnitude of the gravimagnetic field is
given by equations (\ref{8}) and (\ref{8a}) with $\epsilon=1$. Hence, measuring parameter $\epsilon$ is the test of the presence of the {\it extrinsic} gravimagnetic field. This point
has been misinterpreted by Pascual-S\'anchez \cite{pask} who confused
the relativistic effect of the gravimagnetic field caused by the mass current due to the
Lorentz transformation of the gravity field variables with the
measurement of the classic (non-relativistic) R\"omer delay of light emitted by Jupiter.

\subsection{The Lorentz Invariance of Gravity}

The introduction of parameter $\epsilon$, associated with the Lorentz
transformation of gravitational field variables, distinguished between the
gravitational effects due to the fundamental speed of gravity $c_{\rm g}$ from the special relativistic effects caused by the speed
of light $c$. The matrix of the Lorentz transformation of the
gravitational field variables, $\Lambda^\alpha_{\;\beta}(\epsilon)$,
depends on the speed of gravity $c_{\rm g}=c/\epsilon$ which is the same constant as that entering the parametrized Einstein field equations. If $\epsilon\not=1$,
the Lorentz invariance of gravity is broken with respect to the
Lorentz invariance of electromagnetic field and contrariwise. Thus, measuring
the ratio $\epsilon=c/c_{\rm g}$ allows us to test the fundamental compatibility of the gravitational field with the causality principle which is postulated in general relativity but can be violated in alternative theories of gravity \cite{will,mat}.

\subsection{The Aberration and the Speed of Gravity}

The $\epsilon$-parameterization of the Einstein equations with the
single parameter $\epsilon$ helps to keep track of any gravitational
effect associated with the fundamental speed of gravity
$c_{\rm g}=c/\epsilon$.  If $c_{\rm g}=\infty$, all terms in the gravity field equations depending on time derivatives of the
gravitational field would be completely supressed and, hence, propagation of the
gravity force would not be detectable, and the gravimagnetic effect
would be completely unobservable as well.  This interpretation is in a
full agreement with the causal (retarded) nature of gravity which is
revealed in equation (\ref{a5}) describing the null characteristic of
the gravity field equations (\ref{a1}) and (\ref{a2}) connecting the
field point ${\bm x}$ with the retarded position ${\bm z}(s)$ of the
light-ray deflecting body. The null characteristics of gravity must
not be physically confused with the null characteristics of light in
the light deflection experiments. Indeed, light propagates from a
source of light to observer while the null vector ${\bm n}={\bm r}/r$,
where ${\bm r}={\bm x}-{\bm z}(s)$, points from the observer, located
at ${\bm x}$, to the gravitating body located at the point ${\bm
z}(s)$ along the null cone of gravity. In general relativity $c_{\rm g}=c$
and the characteristic hypersurfaces of the null cones for gravity and
light coincide \cite{zakharov}. Nevertheless, null lines of gravity and light
propagation can be distinguished since they are directed to the source
of gravity field (the body) and to the source of light which are
clearly separated in observations \cite{fk,found-phys}.

One notices that in some papers the field is considered as a medium in a flat space-time with a refractive index different from unity 
\cite{pask,skr,mshb,fritt}. This interpretation of the gravity field should be accepted with a great care since the refraction index of any material medium is a scalar and its Lorentz transformation properties are different from those for the gravitational field which is a tensor of second rank. Ignoring this subtlety can lead to misinterpretation of gravitational effects. 

In light-ray deflection experiments the retarded position of the light-ray deflecting body
is measured not by observing body itself, but from precise measurement of the direction of
the gravitational force exerted by the moving body on propagating photons
in the plane of the sky \cite{cqg,k1,fk}. The retardation of gravity
effect can be obtained theoretically in two independent ways: 
\begin{enumerate}
\item by
making use of the retarded Lienard-Wiechert potentials of equations
(\ref{a1})-(\ref{a2}) \cite{ks,k1,found-phys}; 
\item by solving
equations of light propagation in the static frame of the body and,
then, using the Lorentz transformation, as it is done in this paper and
in \cite{klion}.
\end{enumerate}
 The two methods lead to the same results given by
equations (\ref{tdel}) and (\ref{a5}). The Lorentz transformation of
the time delay from a static to a moving frame means that both light
and gravity field variables entering equation (\ref{aa}) must be
transformed simultaneously. The aberration of light transforms vector
$k^\alpha=(1,{\bm k})$ of the light ray alone. If the aberration
of light is taken into account in equation (\ref{aa}) by
transforming the Levi-Civita connection (gravitational force) with $c_{\rm g}\not=c$, the
time delay can not maintain its invariance, and the terms proportional
to $\delta=c/c_{\rm g}-1$ will emerge in equations. If the Lorentz
transformation matrix of the Levi-Civita connection is the same as for
light, then, $\delta=0$ and the time delay is Lorentz-invariant with
the position of the light-ray deflecting body taken at the retarded
time $s$ given by equation (\ref{a5}) with $c_{\rm g}=c$. Thus, the
aberration of light is used in the deflection experiments exclusively
as a calibrating standard for which speed $c$ is fixed and not
measured. The speed of gravity $c_{\rm g}$ is then measured with respect to
this standard \cite{carlip_comment}. It is worth emphasizing that all other gravitational experiments rely exclusively upon the JPL \cite{standish} or equivalent EPM2004 \cite{pit} ephemerides of the solar system constructed in the metric system of units with a numerical value of the speed of light $c=299 792 458$ m$\cdot$s$^{-1}$ precisely (see also \cite{kovs,cons,iau,kaplan1,kaplan2}).

\subsection{The Speed of Gravity, Causality and the Equivalence Principle}\label{cslt}

Review article \cite{will-r} (see also \cite{will-web} and references therein) argues that the speed of propagation of the gravity force from a moving body to the traversing photon is irrelevant in the light-ray deflection experiments done in the Solar system, and that the time delay (\ref{tdel}) and the light-ray deflection angle (\ref{abt}) could be calculated correctly even if the speed of gravity $c_{\rm g}$ would be infinite. In other words, the author of \cite{will-r} maintains the point that the causal structure of the null cone of the gravity field (see Figs. \ref{buh} and \ref{fig_lgwnc}) is not essential to the calculation of the light-ray deflection angle so far as the acceleration of the light-ray deflecting body is ignored. This point of view stems from the belief, explicitly formulated and adopted in the PPN formalism \cite{will}, according to which the principle of equivalence demands nothing about the speed of gravity that determines the hypersurface of causal influence of gravity. We are told that this is because the principle of equivalence operates with derivatives of the metric tensor of the first order (the Christoffel symbols) while the speed of gravity is determined only by the second time derivatives of the metric that appear in the field equations of each metric theory of gravity  (for more detail see section 10.1 in \cite{will}). Therefore, since the Christoffel symbols can be transformed away at any point on the space-time manifold, only the second derivatives (tidal forces) can be locally measured, and only these derivatives show up speed properties of gravity. 

First of all, we notice that the speed of gravity and the (non-local) causal structure of space-time are tightly connected. Gravitational interaction can not propagate information about position of a moving massive body faster than the speed of light if general relativity is correct. Hence, experimental study of the causal properties of the curved space-time manifold answers the question if the speed of gravity has the same numerical value as the speed of light or not. The PPN postulate, mentioned above, assumes that the causality of gravitational field should be relegated only to the gravity field equations. However, the causal structure of gravitational field is derived not so much from the field equations themselves but from the exploration of the behavior of the set of future directed timelike and/or null geodesics in a given space-time manifold \cite{prs,hawe}. This behavior can be determined, at least in a close neigborhood of any event, by the geodesic equations without imposing Einstein's field equations \cite{wald}. The light geodesics define the causal past of observer, that is the region bounded by the past light cone in Fig. \ref{fig_lgwnc}. The causal future of the gravitational field of a massive body, that is the region inside the future gravity cone in Fig. \ref{fig_lgwnc}, is defined by the null geodesics associated with the propagation of gravity field from the body. Causal gravitational interaction of the body with the observed photon implies that the causal future of the gravitational field is not degenerated to a space-like hypersurface that is the photon is gravitationally deflected by the body not at the time of its observation but in the past, so that the observer and the light-ray deflecting body are connected by a null line which must be a solution of the field equations as demonstrated in Figs. \ref{buh} and \ref{fig_lgwnc}. In general relativity, where $c_{\rm g}=c$, the past light cone of observer must touch the future gravity cone along a null line which is a bi-characteristic for the two null cones.

Second, the equivalence principle tells us that in a local reference frame light moves along a straight line \cite{LL,mtw}. But gravitational light-ray deflection experiment is not experiment in a single local frame of reference as photon propagates through a continuous sequence of local frames, thus, accumulating the pointwise influence of the gravitational field at different parts of the light-ray trajectory. Photon's propagator is an integral of the affine connection that is the gravity force shown in the right-hand side of Eq. (\ref{lg}). This propagator contracted with the projection operator $P^{ij}$, yields the integrated deflection angle (\ref{abt}) which can not be transformed away if the space-time is curved, irrespectively of the property of the Christoffel symbols to turn to zero at the origin of each local frame. Gravitational deflection of light is a global phenomenon which is not reduced to the test of the principle of equaivalence. Notice that even in the static gravitational field the principle of equivalence is able to explain only a half of the angle of the gravitational deflection of light while another half of the effect is due to the curvature of space \cite{LL,mtw,will}. In the case when the light-ray deflecting body moves with respect to observer the deflection effect gets more sophisticated as the magnitude and direction of the deflection angle become functions not only of mass but also of the time-dependent position, velocity, etc., of the light-ray deflecting body. Therefore, by precise measurement of the angle of the light deflection (or, equivalently, the time delay) one can determine how strong the gravitational field of the body is, and how fast it interacts with the light ray as it traverses toward observer. If one is able to derive position of the massive body from the precise measurement of the deflection angle (\ref{abt}) and to confirm that the body and the observer are connected by a null line which is a characteristic of the future gravity null cone, it gives a direct proof of the causal nature of the gravitational field and allows us to measure its finite speed of propagation \cite{k1}. However, this measurement of the causal property of gravity is impossible if the gravitational field is static. Indeed, in case of a static planet the gravitational field does not change as photon moves from star to observer, and the causal character of the gravitational field is invisible because the body is always at a fixed distance from the observer (see Fig. \ref{g_l}A). 

The situation changes dramatically if the body moves with respect to observer because it makes the distance between the body and the observer depending on time (see Fig. \ref{g_l}B). In this case, photon traverses through the gravitational field that changes on the light-ray trajectory due to the motion of the body, even if this motion is uniform. Had the speed of the propagation of the body's gravitational field been different from the speed of light it would have unavoidably led to violation of the causal nature of gravity that would be inconsistent with the principle of equivalence as shown in \cite{mat}. For example, the instantaneous propagation of gravity would imply that one could determine current (as opposed to retarded) position of the body on its orbit from observation of the gravitational deflection of light, that is the gravitational field would transmit information about the body's spatial location to observer faster than the light arriving from the star. This violates the principle of causality, and we conclude that correct description of the gravitational physics of the light-ray deflection experiment with a moving body requires taking into account both the light and gravity null cones as demonstrated in Figs. \ref{fig_gwnc}--\ref{g_l}, and supported by calculations in this paper as well as by the discussion given in \cite{cqg}. Additional arguments backing up the concept of the retardation of gravity in the light-ray deflection experiments by a moving body are discussed in our paper \cite{found-phys} both in the framework of general relativity and in a bi-metric theory of gravity proposed by Carlip \cite{carla}. The present paper and calculations given in \cite{found-phys} refute the incomplete understanding of gravitational physics of the light-ray deflection experiments advocated in the PPN formalism \cite{will-r,will-web} and in \cite{carla}. The principle of equivalence does imply the principle of causality for gravitational field and demands the fundamental speed of gravity to be finite (equal to the speed of light in general relativity) which was confirmed in the Solar system experiment of starlight's deflection by the moving Jupiter \cite{fk}.

\subsection{A Graphical Representation of the Aberration of Gravity}

Two methods of measuring the position of a light-ray deflecting body
(gravitational lens) in the sky are: 
\begin{enumerate}
\item[(1)] directly, by observing light
(radio waves) emitted by the body itself. This does not require any measurement of the gravitational light-ray deflection pattern and relies upon propagation properties of the electromagnetic field alone; 
\item[(2)] indirectly, by measuring
the pattern of the gravitational bending of light from stars (radio sources) located in the sky around the lens, and
extrapolating vectors of their gravitational displacement to the
common point of intersection which is the center of mass of the lens (see Fig. \ref{fig1}).
\end{enumerate}
 The second method actually traces the 
lines of the gravitational force exerted on photons by the body. The force lines become apparently measurable in the plane of
the sky due to the gravitational deflection of light. This allows us to map the center-of-gravity of the body, regardless of
whether the body itself emits light or not. This method of \lq\lq
weak-gravitational\rq\rq lensing is used to localize mass-concentration of
dark matter in clusters of galaxies \cite{peter,witt}. In the solar system the weak-lensing method can be effectively used to determine the aberration of the gravitational force caused by motion of a light-ray deflecting body through measuring the retardation of gravity effect that defines position of the body on its orbit \cite{fk} and the magnitude of the gravitational time delay of light, and comparing it
against the retarded position of the body obtained from its direct
radio observations given in the JPL ephemeris \cite{standish}. 

The apparent (optical) and gravitational positions of the light-ray deflecting body from which it deflects light by its gravity force, should
coincide in the frame where the body is static.  However, in a moving
frame, the apparent position of the body is shifted in the plane of the
sky, because of the aberration of light, by the angle $\beta=v/c$. The
gravitational position of the body should be also shifted because of the
aberration of the gravity force lines by the angle $\beta_\epsilon=\epsilon
v/c=v/c_{\rm g}$, as explained in Fig. \ref{fig2}.  If general relativity is
correct, the aberration of gravity force and that of light must be equal and both
apparent and gravitational positions of the body should coincide in any
frame irrespectively of its motion. If the Lorentz invariance of gravity is broken,
the apparent and gravitational positions of the light-ray deflecting body will coincide in a static frame but differentiate in a moving one by the angle $\delta\beta=\beta_\epsilon-\beta$.

Our interpretation of the aberration of gravity arises naturally because the distance ${\bm r}={\bm x}-{\bm z}(s)$ from observer to the gravitating body can be considered as a spatial part of the null vector
$r^\alpha=\bigl(r^0,{\bm r}\bigr)$ with $r^0=c_{\rm g}(t-s)$, where $t$ is time
of observation, and $s$ is retarded time defined by equation
(\ref{a5}). The null vector $r^\alpha$ transforms in accordance with
the group of the Lorentz transformation of gravitational field
described by the matrix $\Lambda^\alpha_{\;\beta}(\epsilon)$ and its
direction defines the null direction of the propagation of gravitational force from the light-ray deflecting body
to the light particle. This null direction of the gravity force propagation changes (aberrates) when one goes from one frame to another. The magnitude of the aberration of gravity effect is linear with respect
to $v/c_{\rm g}$ like the aberration of light is linear with respect to the ratio of
$v/c$.

This reasoning makes it evident that the measurement of the aberration of gravity can be obtained by
measuring the deflection of light (radio waves) in a frame that is not
stationary with respect to the light-ray deflecting body.  Our 2002
Jupiter deflection experiment \cite{fk} was specifically designed
to measure the speed of gravity force by observing the magnitude of its aberration via excess to the gravitational Shapiro time delay caused by retardation in propagation of gravity from Jupiter to radio photons from quasar (see text
below).  Future optical deflection observations by space missions SIM \cite{sim} or
Gaia \cite{gaia} of stars near the limb of Jupiter or Saturn, should
bring about more accurate results. Combining VLBA and Square Kilometer Array \cite{ska} to the ineterferometer with inter-continental baseline can be also used to conduct this type of the gravitational experiments with unparalleled degree of precision \cite{freid}.

The Sun is another moving body with
respect to the earth as observed from the geocentric frame, but only radio observations of radio sources
near the Sun that can reach accuracy being sufficient to measure the aberration of gravity effect, are currently feasible (see \cite{kwtn} for future prospects).  Previous solar bending observations have
been always analyzed in the barycentric frame (with respect to which the Sun
is almost static so that the gravity aberrational term is negligibly
small) because the main goal was the precise measurement of the PPN
parameter $\gamma$ \cite{fsr,leb}.  However, the analysis of the
experiments directly in the geocentric frame, where the Sun moves with velocity
30 km/s, can determine not only the PPN parameter $\gamma$, but the
gravity aberration term proportional to $\delta\alpha(\epsilon)$,
derived below and shown in equation (\ref{ura2}). This is because the Lorentz transformation from the barycentric to geocentric frame probes transformation properties of the gravitational force which are controlled by the speed of gravity $c_{\rm g}$.

\section{Experimental VLBI Measurement of the Aberration of Gravity}

\subsection{The Magnitude of the Aberration}

The formulae for the aberration term can be determined from equations
(\ref{a5}) and (\ref{abt}). Let us assume that the impact distance $d$
of the light ray to the massive body is small ($d\ll r$) so that we can expand
equation (\ref{abt}) in a Taylor series with respect to the parameter
$d/r$. Then, we obtain \cite{km,ks}
\begin{equation}
\label{hkx}
r-{\bm k}\cdot{\bm r}=\frac{d^2(\epsilon)}{2r}+O\left(\frac{d^4}{r^3}\right)\;,
\end{equation}
where $d(\epsilon)=|{\bm\xi}(\epsilon)|$ is the impact distance of the
light ray with respect to the retarded position ${\bm
z}(s)$ of the body, ${\bm \xi}(\epsilon)={\bm
k}\times[{\bm r}(s)\times{\bm k}]$, ${\bm
r}(s)={\bm x}-{\bm z}(s)$,
$r(s)=|{\bm r}(s)|$, ${\bm x}$ is the
position of observer at the time of observation $t$, the
retarded time $s=s(\epsilon)$ is calculated from equation (\ref{a5}) that, after accounting for definition $c_{\rm g}=c/\epsilon$, acquires the following form
\begin{equation}
\label{qws2}
s=t-\epsilon\,\frac{r(s)}{c}\;.
\end{equation} 
Using equation (\ref{hkx}) and neglecting the explicit
velocity-dependent terms, we can simplify equation (\ref{abt}) for the
total angle of the light deflection and reduce it to the form
\begin{equation}
\label{qws}
{\bm\alpha}(\epsilon)=2(1+\gamma)\frac{GM}{c^2 }\frac{{\bm\xi}(\epsilon)}{d^2(\epsilon)}\;,
\end{equation}
where the gravitational deflection of light vector ${\bm\alpha}\equiv\left(\alpha^i\right)$, and $M$ is mass of the body.

If the speed of gravity $c_{\rm g}=c/\epsilon$ is different from the speed
of light $c$ ($\epsilon\not=1$) the Lorentz-invariance of gravity is
broken and the aberration of gravity differs from that of
starlight. This would produce a small tangential (in the plane of the
sky) displacement
\begin{equation}
\label{qk5}
\delta{\bm\alpha}(\epsilon)={\bm\alpha}(\epsilon)-{\bm\alpha}
\end{equation} 
of star's position, deflected by the body as measured in the moving
frame, compared with that ${\bm\alpha}\equiv{\bm\alpha}(\epsilon=1)$ expected from general relativity.  The approximate position and motion of the body,
however, must be known in order to estimate an accurate value for
${\bm\alpha}$ which is taken as a reference in measuring $\epsilon=c/c_{\rm g}$. For solar system
objects, exceedingly accurate ephemerides are available \cite{standish,pit}, so that ${\bm\alpha}$ can be predicted with an accuracy being sufficient to set a stringent upper limit on the Lorentz
violation of gravity field and its fundamental speed $c_{\rm g}$, which is not limited by
the uncertainty in position and/or velocity of the body.

The aberration of gravity effect is estimated after expansion of
${\bm\alpha}(\epsilon)$ in a Taylor series around $\epsilon=1$ and making use of
equations (\ref{qws2}) and (\ref{qk5}). One has (see Fig. \ref{fig3})
\begin{equation}
\label{qk6}
{\bm\alpha}(\epsilon)={\bm\alpha}+\left[\frac{\partial{\bm\alpha}(\epsilon)}{\partial
\epsilon}\right]_{\epsilon=1}(\epsilon-1)+O\left[(\epsilon-1)^2\right]\;,
\end{equation}
that after substitution to equation (\ref{qk5}) and taking the derivative yields
\begin{eqnarray}
\label{qk7}
\delta{\bm\alpha}(\epsilon)&=&\frac{\alpha}{\theta}\,\delta{\bm\beta}(\epsilon)\;,\\
\label{qk7a}
\alpha&=&2(1+\gamma)\frac{GM}{c^2 d}\;,\\
\label{qk7b}
\theta&=&\frac{d}{r}\;,\\
\label{qk7c}
\delta{\bm\beta}(\epsilon)&=&{\bm\beta}_\epsilon-{\bm\beta}=(\epsilon-1)\frac{{\bm v}_T}{c}\;,
\end{eqnarray}
where ${\bm v}_T={\bm k}\times({\bm v}\times{\bm k})$ is the transversal velocity of the light-ray deflecting body in
the plane of the sky and all values of the functions entering
equations (\ref{qk7})--(\ref{qk7c}) are taken at the retarded instant
of time for $\epsilon=1$.  Clearly, the equation (\ref{qk7c}) compares the aberration of gravity, ${\bm\beta}_\epsilon={\bm v}_T/c_{\rm g}$, with respect to the aberration of light, ${\bm\beta}={\bm v}_T/c$ which are both linear with
respect to body's velocity ${\bm v}$ but have different physical origin associated with the transformation properties of the gravitational and electromagentic field. Measuring presumable tangential displacements $\delta{\bm\alpha}$ in the gravitational deflection of starlight sets a stringent limit on the aberration of gravity ${\bm\beta}_\epsilon$ and, hence, on a possible discrepancy between the speed of gravity $c_{\rm g}$ and that of light $c$.

Equation (\ref{qk7}) was used in our papers \cite{k1,fk} to estimate the
aberration of the gravity force in the light bending experiment
with Jupiter as the gravitational lens. We analyzed the data in the
barycentric frame of the solar system where Jupiter has an orbital
speed of $4.5\times 10^{-5}~c$. The maximal magnitude of the
aberration of gravity force caused by the motion of Jupiter reaches
$\delta\alpha\simeq 10$ mas when Jupiter is at the distance 6
astronomical units (6 AU) from the earth and light passes near
Jupiter's limb.  For the 2002 experiment, the quasar passed $3.7'$
from Jupiter and the deflection of light caused by the aberration of gravity force was 0.05 mas. It was measured
to an accuracy of 20\% \cite{fk}, thus, disproving experimentally the erroneous statements by T. van Flandern \cite{vfl1,vfl2} about the nature of gravity and its speed of propagation. 

Most of the previous radio gravitational bending experiments occurred in
early October when the Sun passed in the plane of the sky closely in front of the strong quasar 3C279.  The
aberration of gravity term as viewed in the Shapiro time delay computed in the earth-centered frame is relatively
large since the Sun moves with a $v_\odot\simeq 30$ km/s in this
frame.  The estimate of the tangential component in the gravitational bending of light caused by the aberration of gravity relative to the
aberration of light effect, yields
\begin{equation}
\label{ura}
\delta\alpha(\epsilon)=\alpha_\odot\left(\frac{1+\gamma}{2}\right)\left(\frac{A}{R_\odot}\right)\left(\frac{R_\odot}{d}
\right)^2\,\delta\beta(\epsilon)\;,
\end{equation}
where $\alpha_\odot=1.75''$ is the light deflection on the solar limb,
$A\equiv 1$ {\rm AU} is one astronomical unit, $R_\odot=7\times
10^{10}$ cm is radius of the Sun. Substituting in the numbers one
obtains
\begin{equation}
\label{ura2}
\delta\alpha(\epsilon)=37.5 (\epsilon-1)\left(\frac{1+\gamma}{2}\right) \left(\frac{R_\odot}{d}\right)^2 \mbox{mas}\;. 
\end{equation}
If we assume the minimal value of the impact parameter that is
permitted in radio observations by the solar corona as $d\simeq 4 R_\odot$, the maximal
value of the aberration of gravity effect (see Figure \ref{fig3}) is
$\delta\alpha\simeq 2344$ $\mu$arcsec for $\epsilon=0$. Assuming that
the precision of VLBI measurement of the quasar's position is 20
$\mu$arcseconds we shall be able to measure the aberration of gravity
effect in the solar light-ray deflection observations with the accuracy approaching, in principle, to 1\% as contrasted to 20\% in
the case of the Jovian deflection experiment. 

\subsection {Lorentz Transform Between Frames}

The fundamental measurable of VLBI is the difference in the arrival
time of radio waves from an external source impinging on each
telescope (station) in the array. This time difference is inferred by
the phase difference of the two radio waves.  The source of radio
waves can be a quasar at cosmological distance or a spacecraft in the
solar system.  The former produces plane waves and is stationary in
the sky, whereas the latter is in the solar system which adds minor
complications to the analysis that are not considered here.

Let us consider two frames, the barycentric frame of the solar system
and the geocentric frame, on which VLBI array analyses are done.  As a
quasar signal passes by a radio-wave deflecting body (Sun, planet) on
its way to the array, the major components of the observed delay
(difference in the arrival time between two stations, as measured the
phase difference of the two signals) in the barycentric frame are given by
equation (\ref{tdel})
\begin{eqnarray}
\label{ga0}
T_2-T_1&=&-\frac{1}c{\bm K}\cdot{\bm B}+H(T,{\bm X})+E(T,{\bm X},\nu)\;,\\
\label{g2}
H(T,{\bm X})&=&-(1+\gamma)\left(1-\frac{1}{c_{\rm g}}{\bm K}\cdot{\bm V}\right)\frac{GM}{c^3}\ln\left(\frac{R_1-{\bm K}\cdot{\bm R}_1}{R_2-{\bm K}\cdot{\bm R}_2}\right)\;,\\\label{g3}
{\bm R}_i&=&{\bm X}_i-{\bm Z}(S_i)\;,\qquad(i=1,2),\\
\label{g4}
S_i&=&T_i-\frac{R_i}{c_{\rm g}}\;,\qquad(i=1,2).
\end{eqnarray} 
where we use capital letters $T$ and ${\bm X}$ to denote the time and
spatial coordinates in this frame, $T_i$ ($i$=1,2) is the time of
arrival of radio wave to $i$-th VLBI stations, ${\bm K}$ is the unit
vector toward the radio source, $H$ is the gravitational time delay,
$E$ represents all additional sources of delay error and is (radio) frequency
$\nu$ dependent (egs. solar coronal refraction), ${\bm X}_i={\bm
X}(T_i)$ is coordinate of $i$-th VLBI station at the time $T_i$, with
${\bm B}={\bm X}_2-{\bm X}_1$ as the baseline between the two VLBI
stations, ${\bm Z}(S_i)$ is a retarded coordinate of the light-ray
deflecting body at the retarded time $S_i$, ${\bm V}(S)=d{\bm
Z}(S)/dS$ is the barycentric velocity of the body taken at the retarded time, and ${\bm R_i}$ is
the distance between the station $i$ and the retarded position of the light-ray deflecting
body.  In \S {VIB.5} we will outline the methods that are used to
isolate the gravitational delay, $H$, by removing the delay errors
$E$. The solar light-deflection experiments neglect the slow motion of
the Sun with respect to the barycenter, so that its coordinate ${\bm
Z}$ is considered as fixed during the time of the experiment. The
static-field relativistic correction $H$ to the time delay was derived
by \cite{shapiro}.  Note that the static-field approximation can not
be applied in the deflection experiment of Jupiter in the barycenter
frame for it moves with respect to this frame rather rapidly; hence, the aberration of gravity effect from its velocity
in this frame was taken into account in the analysis of the September
2002 experiment with Jupiter \cite{k1,fk}.

The Lorentz transformation of the electromagnetic and gravitational
fields to a frame moving with velocity ${\bm v}$ with respect to the
barycentric frame, using equation (\ref{tdel}), gives the following
components for the gravitational delay in the moving frame
\begin{eqnarray}
\label{ga1}
t_2-t_1&=&-\frac{1}c{\bm k}\cdot{\bm b}+h(t,{\bm x})+e(t,{\bm x},\nu')\;,\\
\label{g6}
h(t,{\bm
x})&=&-(1+\gamma)\left(1-\frac{1}{c_{\rm g}}{\bm k}\cdot{\bm v}\right)\frac{GM}{c^3}\ln\left(\frac{r_1-{\bm
k}\cdot{\bm r}_1}{r_2-{\bm k}\cdot{\bm r}_2}\right)\;,\\\label{g7}
{\bm r}_i&=&{\bm x}_i-{\bm z}(s_i)\;,\qquad(i=1,2),\\\label{g8}
s_i&=&t_i-\frac{r_i}{c_{\rm g}}\;,\qquad(i=1,2).
\end{eqnarray}
where $t$ and ${\bm x}$ denote the time and spatial coordinates in
this moving frame, $t_i$ ($i$=1,2) is the time of arrival of radio
wave to $i$-th VLBI stations, ${\bm k}$ is the unit vector toward the
radio source, $h$ is the gravitational time delay, $e$ represents all
additional sources of delay error and is frequency $\nu'$  dependent (notice that $\nu'\not=\nu$ because of the Lorentz transformation), ${\bm x}_i={\bm x}(t_i)$ is
coordinate of $i$-th VLBI station at the time $t_i$, with ${\bm
b}={\bm x}_2-{\bm x}_1$ as the baseline between the two VLBI stations.
In converting to the moving frame, the speed of gravity, $c_{\rm g}$, appears as a consequence of transformation of the gravitational field variables in
two places: first, in the aberrational term ${\bm k}\cdot{\bm v}/c_{\rm g}$ in front of the logarithm;
and second, in the retarded term ${\bm r_i}$, the distance between the
station $i$ and the light-ray deflecting body taken at the retarded
time $s_i$.

The general form of the time delay equation which takes into account
not only linear but quadratic and high-order velocity-dependent
corrections as well, has been derived in \cite{km,ks} and \cite{klion}
under assumption that $c_{\rm g}=c$.

\subsection{Measuring the PPN Parameter $\gamma$}

The gravitational time delay is proportional to $(1+\gamma)$, where the
parameter $\gamma$ is a static measure of space curvature and is unity in general relativity and zero in the
Newtonian gravity \cite{will}. The quantity $(\gamma-1)$ measures the
degree to which gravity is not a pure geometric phenomenon and other
gravity-generating scalar fields are involved \cite{kv}. These fields might
dominate in the early Universe, but would have now weakened so as to
produce tiny, but presumably detectable effects \cite{DN,kv}.

Although all VLBI experiments are done on or near the earth, most
astrometric quantities, including $\gamma$, are more conveniently
analyzed in the solar system barycentric reference frame $(T,{\bm X})$
because most velocity-dependent relativistic corrections are zero in
this frame.  Hence, VLBI reduction systems use this frame \cite{cons,iau,kaplan1,kaplan2}
where the gravitational time delay is given by equation (\ref{g2})
with ${\bm V}=0$, since this velocity is very small.  Thus, in the
solar light-ray deflection experiments, we should analyze the measurements of the
gravitational delay as a function of the Sun-quasar angular separation
in the barycentric frame to determine $\gamma$. The estimated accuracy
in measuring $\gamma-1$ each observing day for this kind of experiments with 3C279 that passes near the Sun each October,
is given in Table 1 and is expected to be about $7\times
10^{-5}$. The best previous measurement of $\gamma=2.3\times 10^{-5}$
was achieved during the conjunction of the Sun and Cassini spacecraft
in 2002 using the gravitational-induced frequency change of the
spacecraft transmitters as Cassini went behind the Sun on its way to
Saturn \cite{bert}.

\subsection{Measuring the Aberration of Gravity and Its Fundamental Speed}

In the 2002 Jupiter deflection experiment, the aberration of gravity in the retarded position of Jupiter
was measured in the barycentric frame by making use of equations
(\ref{ga0})--(\ref{g4}) because Jupiter moves with respect to the
barycenter of the solar system \cite{fk}.  But, the barycentric
velocity of the Sun is rather small, so its gravity-field aberration in the barycentric frame is not easy detectable.  However, the aberration of the solar gravity field, is
measurable from the earth-center frame since the deflecting object,
the Sun, is moving about 30 km/s with respect to this frame and its gravitational field contains the gravimagnetic component that drags photons in the direction of its motion \cite{pla}.

The experimental observable is the measured delay difference of the
electromagnetic wave between two stations.  This delay (technically,
it is a phase difference, but can be converted into delay) is a scalar
function of time and baseline. In the absence of gravity field this
delay is Lorentz invariant, that is \cite{found-phys,kozer,kopsva}
\begin{equation}
\label{vok}
T_2-T_1+\frac{1}c{\bm K}\cdot{\bm B}=t_2-t_1+\frac{1}c{\bm k}\cdot{\bm b}\;.
\end{equation}
Therefore, the gravitational part of the time delay difference must be also
{\it independent} of the observing frame used in the analysis (because it is a scalar),
\begin{equation}
\label{deldif}
H(T,{\bm X})=h(t,{\bm x})\;,
\end{equation}
where the conversion of all relevant quantities from the
barycentric to the geocentric frame must be done precisely.

In the following description of the conversion of the reference frames,
we will denote the barycentric coordinates of the Sun ${\bm Z}_\odot$,
those of the geocenter ${\bm Z}_\oplus$, and the geocentric coordinates
and velocity of the Sun are ${\bm z}_\odot={\bm Z}_\odot-{\bm
Z}_\oplus$ and ${\bm v}_\odot={\bm V}_\odot-{\bm V}_\oplus$ respectively. Relativistic
corrections to these definitions are quadratic with respect to $v/c$
\cite{kv,iau,kaplan1} and can be neglected in current practical application of
equation (\ref{deldif}). The barycentric coordinates and velocity of
the Sun and the geocenter are also known as functions of the
barycentric time $T$ from accurate solar system ephemerides
\cite{standish,pit}. The barycentric time $T$ is related to the geocentric
time $t$ measured on the earth by a time transformation in which the
difference is quadratic with respect to $v/c$ terms \cite{brk,fu}, and,
thus, negligibly small in dealing with equation (\ref{deldif}).  The
Lorentz transformations needed for the frame conversions in equation
(\ref{deldif}) are as follows.

First, the two unit vectors ${\bm K}$ and ${\bm k}$, characterizing
the direction of the propagation of light in two different frames, are
connected by the Lorentz transformation equation (\ref{pox}) for
electromagnetic field. In the linear approximation this transformation
is reduced to the classic expression for the aberration of starlight
\begin{equation}
\label{g5}
{\bm k}={\bm K} -\frac{1}{c}\;{\bm K}\times\left({\bm V}_\oplus\times{\bm
K}\right)+O\left(\frac{v^2}{c^2}\right)\;.
\end{equation}
Barycentric earth's velocity ${\bm V}_\oplus$ and the astrometric coordinates of the vector ${\bm K}$ for
3C279 and the other calibrators are accurately known, so the
coordinates of the unit vector ${\bm k}$ in the geocentric frame can be
easily calculated from equation (\ref{g5}), and vice versa.

Second, the retarded geocentric coordinate of the moving Sun, ${\bm
z}_\odot(s_i)$, entering equation (\ref{g6}), can be represented in
the first approximation as a difference between the barycentric
coordinates of the Sun, ${\bm Z}_\odot$, and the geocenter, ${\bm
Z}_{\oplus}$,
\begin{equation}
\label{g9}
{\bm z}_\odot(s_i)={\bm Z}_\odot(s_i)-{\bm Z}_{\oplus}(s_i)+O\left(\frac{v^2}{c^2}\right)\;,
\end{equation}
where the retarded time 
\begin{equation}
\label{grum}
s_i=t_i-\frac{1}{c_{\rm g}}|{\bm x}(t_i)-{\bm z}_\odot(s_i)|\;.
\end{equation}
Next, the coordinates of the VLBI station in the moving and static frames are
related by equation
\begin{equation}
\label{gkt}
{\bm x}_i={\bm X}_i(t_i)-{\bm Z}_{\oplus}(t_i)+O\left(\frac{v^2}{c^2}\right)\;,
\end{equation}
taken at the time of observation $t_i$. The retardation in the
barycentric coordinate of the center of mass of the Sun ${\bm
Z}_\odot(s_i)$ does not affect its value significantly because it
moves very slowly around the barycenter of the solar system; hence
equalities ${\bm Z}_\odot(s_i)={\bm Z}_\odot(t_i)$, ${\bm
V}_\odot(s_i)={\bm V}_\odot(t_i)$ can be used, although the precise
correction for the retardation is made in the analysis software to eliminate any possible error.  On the other
hand, the orbital motion of the earth around the barycenter of the
solar system is significant and the effect of the retardation of the
force of gravity (as it propagates in the geocentric frame from the Sun to the earth) in the
coordinates of the geocenter must be taken into account. It yields for
the retarded barycentric coordinate of the geocenter,
\begin{eqnarray}
\label{asp}
{\bm Z}_{\oplus}(s_i)&=&{\bm Z}_{\oplus}(t_i)+{\bm V}_{\oplus}(t_i)(s_i-t_i)+O\left[(s_i-t_i)^2\right]\\\nonumber&=&
{\bm Z}_{\oplus}(t_i)-\frac{1}{c_{\rm g}}{\bm V}_{\oplus}(t_i)|{\bm x}_i-{\bm z}_\odot(s_i)|+O\left(\frac{v^2}{c^2}\right)\;,
\end{eqnarray}
where the retardation of gravity correction proportional to $s_i-t_i$, is linear with respect to
the orbital velocity of the earth and has the same order of magnitude as
that for the aberration of light given in equation (\ref{g5}),
if $c_{\rm g}=c$. Finally, transformation of the spatial coordinates from
equation (\ref{g7}) to (\ref{g3}) becomes
\begin{equation}
\label{frp}
{\bm r}_i={\bm R}_i-\frac{1}{c_{\rm g}}{\bm V}_\oplus R_i+O\left(\frac{v^2}{c^2}\right)\;.
\end{equation}
Introducing unit vectors ${\bm n}_i={\bm r}_i/r_i$ and  ${\bm N}_i={\bm R}_i/R_i$ we can re-write equation (\ref{frp}) to the following form
\begin{equation}
\label{fqx}
{\bm n}_i={\bm N}_i -\frac{1}{c_{\rm g}}\;{\bm N}_i\times\left({\bm V}_\oplus\times{\bm
N}_i\right)+O\left(\frac{v^2}{c^2}\right)\;.
\end{equation}

Note, that the transformation of the retarded coordinates of the earth and the
the Sun relates exclusively to the transformation of the gravitational
field variables, while the transformation of electromagnetic field (radio wave from the quasar) is associated
with the change (\ref{g5}) in the direction of its propagation. Thus,
equation (\ref{fqx}) describes the aberration of gravity, while
equation (\ref{g5}) is the aberration of starlight (see Fig. \ref{fig2}).  These aberrations
could be different if the fundamental speed of gravity $c_{\rm g}$ were not
equal to the speed of light $c$.  This difference would lead to the
gravitational deflection of light coming from a quasar, different than
the prediction of general relativity as shown in Fig. \ref{fig3}.

The velocity-dependent term in front of the logarithmic function in
equation (\ref{g2}) is proportional to the radial velocity of the Sun
with respect to the barycenter, ${\bm K}\cdot{\bm V}_\odot$. This
velocity is too small and its effect on the time delay $H$ in the
barycentric frame is not currently detectable. On the other hand, the
radial velocity-dependent term in front of the logarithm in equation
(\ref{g6}) is ${\bm k}\cdot\left({\bm V}_\odot-{\bm
V}_\oplus\right)$. It includes the orbital velocity ${\bm V}_\oplus$
of the geocenter and, in principle, could be measured.  However, the
orbital velocity of the geocenter is projected on the direction to the
source of light ${\bm k}$, and reduces its effect on the time delay to
the size of about 1 $\mu$arcsecond which is somewhat too small for
current VLBI technology.  Therefore, we shall neglect the
velocity-dependent terms in front of the logarithmic functions.

After the Lorentz transforms are made from the barycenter to the
earth-center frame, we can measure the fundamental
speed of gravity $c_{\rm g}$ by solving equations (\ref{ga1})--(\ref{g8})
for the gravity Lorentz-invariance violating parameter $\epsilon$.  An estimate
of its measured accuracy from the solar deflection
experiment is a few percent and depends on how accurately the
coronal refraction is removed.

We emphasize that the above transformations must be made as accurately
as possible, given the desired accuracy of the observations.  For
example, the unperturbed light particle moves in vacuum with constant velocity $c$ in any
frame.  Since time can be measured presently with much better accuracy
than space intervals, this led to abandoning the measurement of the
speed of light $c$ in vacuum more and more precisely. Thus, the numerical
value of the speed of light has been fixed $c=299 792 458$
m$\cdot$s$^{-1}$, and it is this value which is used to measure
distances in the solar system.  The most accurate barycentric
coordinates of the solar system bodies -- Sun, planets, and their
satellites -- are normalized to this numerical value of the speed of
light $c$ and are tabulated in the solar system ephemeris \cite{standish,pit}.
We should adopt this metric system of units and consider the speed of light
$c$ as known parameter that is not subject to measurement. This point
is of paramount importance for unambiguous physical interpretations of gravitational
experiments in the solar system and beyond \cite{carlip_comment}. 

\ack
We wish to acknowledge the grant support of the University of
Missouri-Columbia and the Eppley Foundation for Research. We thank anonimous referee for constructive suggestions and comments which helped us to improve the manuscript. The
National Radio Astronomy Observatory is a facility of the National
Science Foundation, operated under cooperative agreement by Associated
Universities, Inc. 

\newpage
\begin{figure*}
\includegraphics[height=10cm]{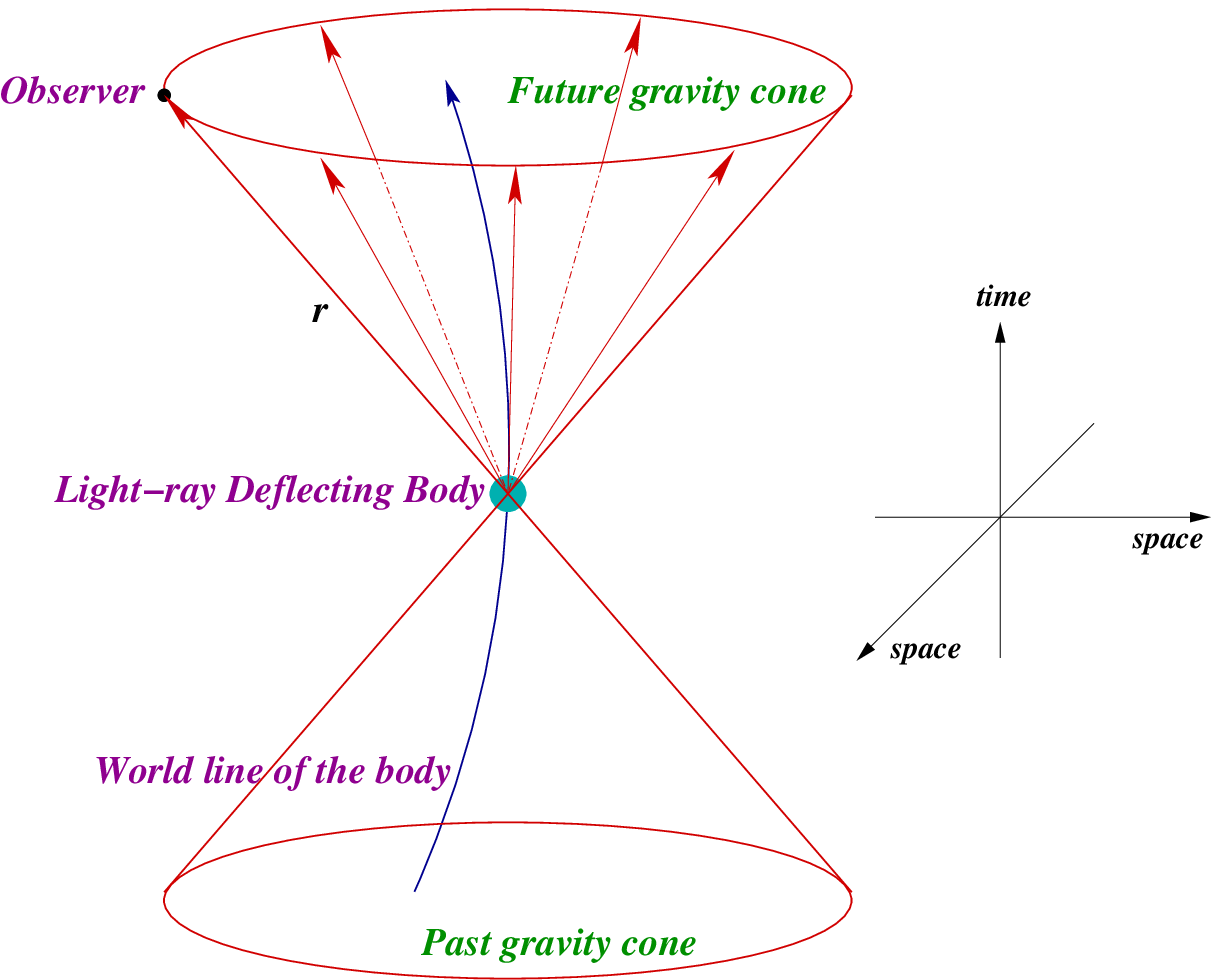}
\caption[Gravity null cone]{\label{fig_gwnc}
Gravitational field of a moving body is a retarded solution of the gravity-field wave equation (\ref{a1}), (\ref{a2}). In general relativity, gravitational field originates at the body location at each instant of time $s$ on its world line at the point ${\bm z}(s)$, and progresses on the hypersurface of a null cone, attached by its vertex to the point ${\bm z}(s)$, from the past to the future. Directions of the propagation restricting the domain of the causal influence of the gravitational field of the body located at the point ${\bm z}(s)$, are shown by arrows. The conventional interpretation of this drawing is that the moving body "emits" the gravitational field at time $s$, and observer measures the gravitational field at time $t$, when the body is located at the retarded position ${\bm z}(s)$ on its orbit at the retarded time $s=t-r/c_{\rm g}$, which is a null characteristic of the gravitational field as follows from equation (\ref{apom}). For simplicity, the picture shows only one null cone of gravity field. Effectively, the entire sequence of null cones must be shown for each instant of time corresponding different positions of the body on its world line.}
\end{figure*}
\newpage
\begin{figure*}
\centerline{\includegraphics[height=8cm]{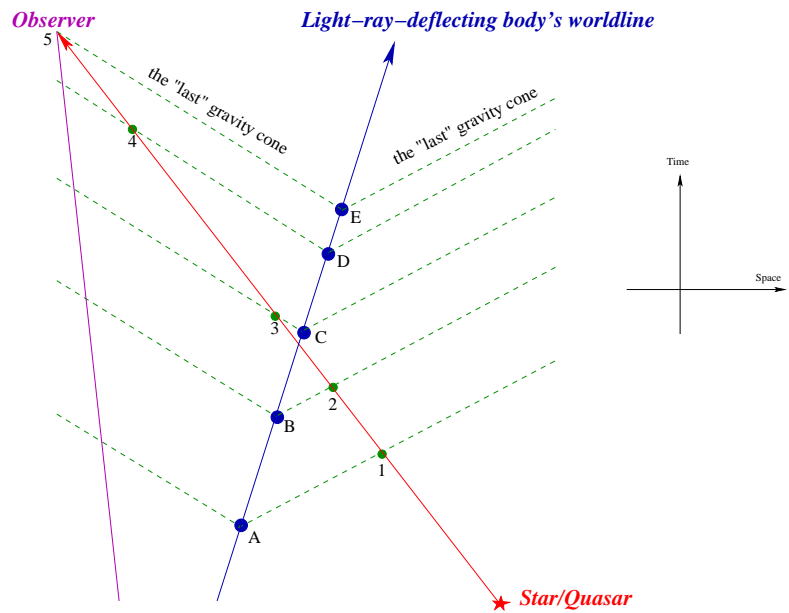}}
\caption[Sequence of gravity null cones interacting with light]{\label{buh}
A photon is emitted by a source of light (star/quasar) and propagates along a hypersurface of light null cone (solid red line) toward observer. Gravitational field of a light-ray deflecting body propagates along the gravity null cone (dashed green lines) and interacts with the photon at the points of intersection of the light and gravity null cones (points 1 through 5). Gravitational light-ray deflection angle and time delay are integrals along the light null cone but their numerical values are determined by the retarded position of the body on the gravity null cone connecting the body and the photon as follows from equations (\ref{tdel}) and (\ref{abt}). Observer measures the time delay and the deflection angle of the photon at time $t$ corresponding to point 5. At this time the time delay and the deflection angle are determined by the body position taken on the "last" gravity null cone at point E. Notice that the point of the closest approach of light ray and/or the time of the clossts approach of the photon to the body are irrelevant to the process of interaction of light with gravitational field of a moving body, as follows from the result of integration of the light-ray propagation equation. This picture illustrates the case when the gravity and light null cones are formally different ($c_{\rm g}\not= c$). In general relativity the gravity null cone coincides with the light null cone and the photon, after it passes the body, moves in the gravitational field of the body which is practically "frozen" for this photon \cite{ks} because the speed of gravity is the same as the speed of light.} 
\end{figure*}
\newpage
\begin{figure*}
\includegraphics[height=8cm]{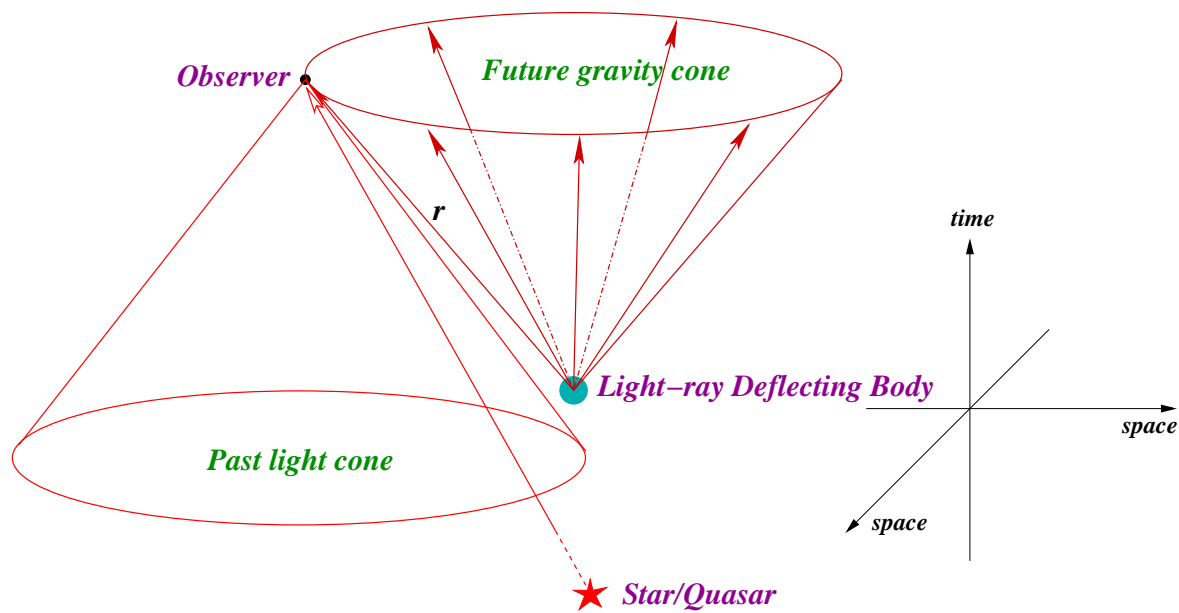}
\caption[Interacting gravity and light null cones]{\label{fig_lgwnc}
Gravitational field of a moving body affects only the particles lying on the hypersurface of the future gravity null cone due to the gravity causal nature. A photon emited by a source of light (star, quasar) at time $t_0$ arrives to observer at time $t$ along a null direction of the past light cone with the observer located at its vertex. As explained in section 5 and in caption to Fig. \ref{buh}, the photon detected at time $t$, is deflected by planet's gravity force from the planet's retarded position taken at time $s=t-r/c_{\rm g}$  that is a null characteristic of the retarded solution of the gravity-field wave equations (\ref{a1}), (\ref{a2}). This effect of the retardation of gravity can be observed by measuring the amount and direction of the gravitational deflection of starlight by a moving body.
In general relativity $c_{\rm g}=c$, hence, the future gravity null cone of the body and the past light cone of the observer must coincide along the null direction which is simultaneously characteristic of the Maxwell and Einstein equations. This bi-characteristic nature of general relativity led to the confusing statements in literature (see \cite{will-r,will-web} and references therein) that the Jovian light-ray deflection experiment measured the speed of light coming from the quasar. In fact, the light was used to measure the amount and direction of its gravitational deflection from which the orbital position of the Jupiter was derived as predicted by the equations (\ref{tdel}), (\ref{a5}) \cite{fk}. }
\end{figure*}
\newpage
\begin{figure*}
\includegraphics[height=10cm]{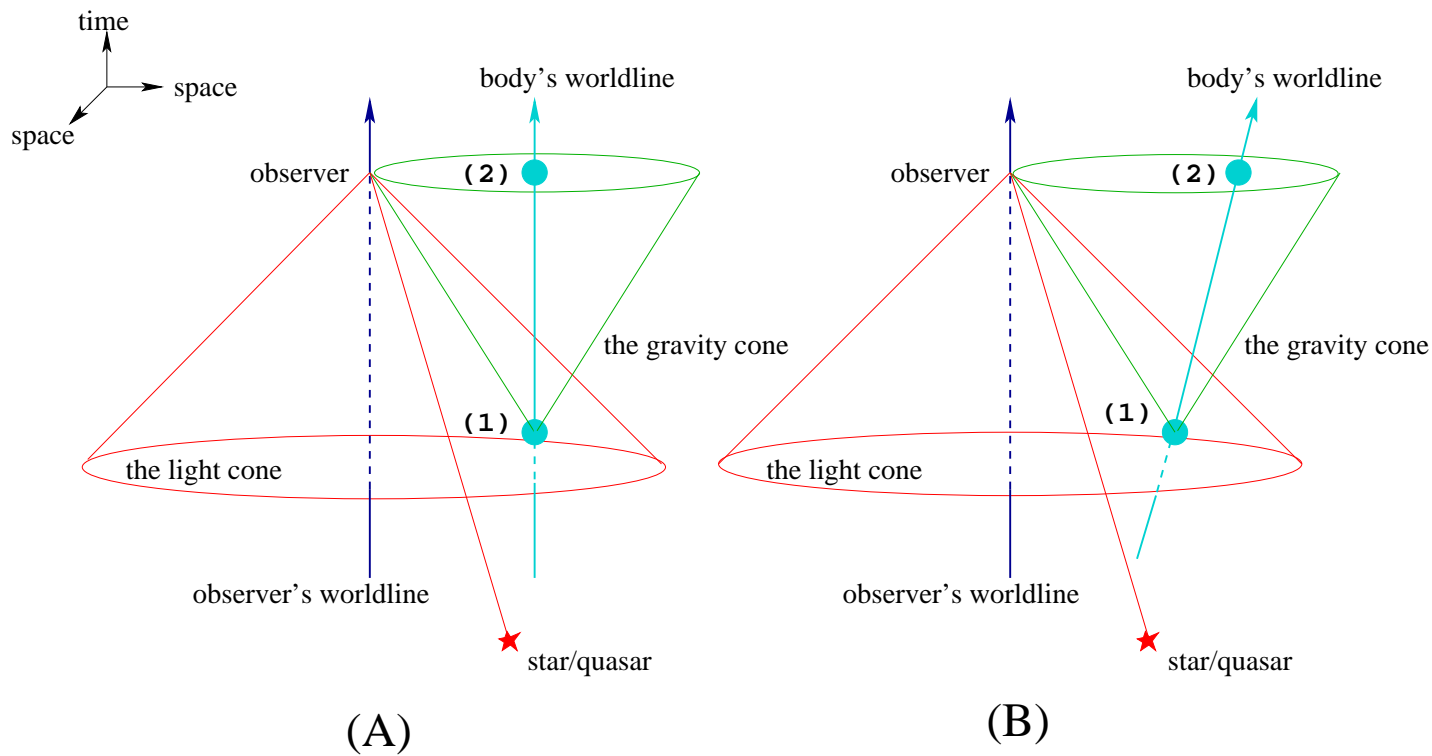}
\caption[Static versus dynamic light-ray deflection]{\label{g_l}
Light-ray deflection by a static (A) and moving (B) body. In case (A) the spatial distance between the body and observer does not change as light propagates. Thus, measuring the deflection of light does not allow us to determine experimentally whether the gravity force of the body acts with retardation from position (1), or instantaneously from position (2). In case (B) the spatial distance between the body and the observer varies as the photon travels toward observer. The retarded interaction of gravity with light becomes apparent since measuring the angle of the gravitational deflection of light allows us to distinguish between positions (1) and (2) of the body on its world-line making the causal structure of the gravity cone clearly discernible and measurable. This kind of gravitational experiments works perfect even if the body moves uniformly with constant velocity with respect to observer so that the gravitational radiation is not emited.}  
\end{figure*}

\clearpage
\begin{figure*}
\includegraphics[height=18cm]{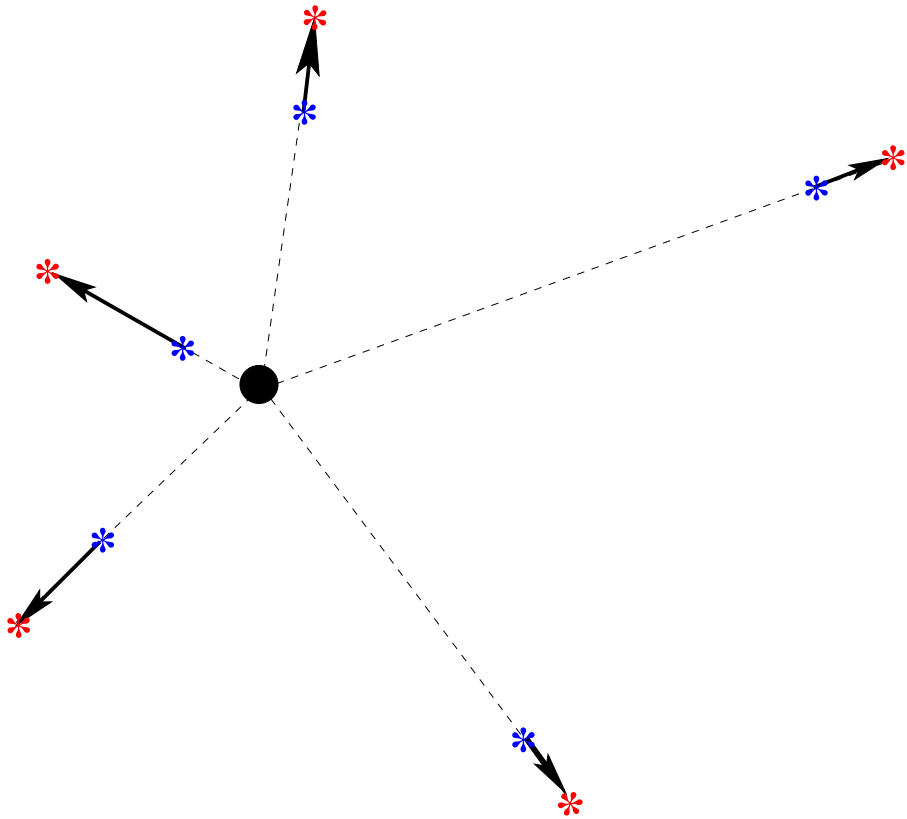}
\caption[Light deflection in the static frame]{Deflection of starlight
caused by spherical gravitational field of a body. Arrows show the angular displacement of stars from their undeflected positions and map the lines of the gravity force deflecting the starlight. The point of intersection of these lines defines the
body's center of gravity in the sky.  
Apparent position of the body visible in optics/radio and that found from the gravitational
deflection of starlight coincide in the static frame.}
\label{fig1}
\end{figure*}
\clearpage
\begin{figure*}
\includegraphics[height=21cm]{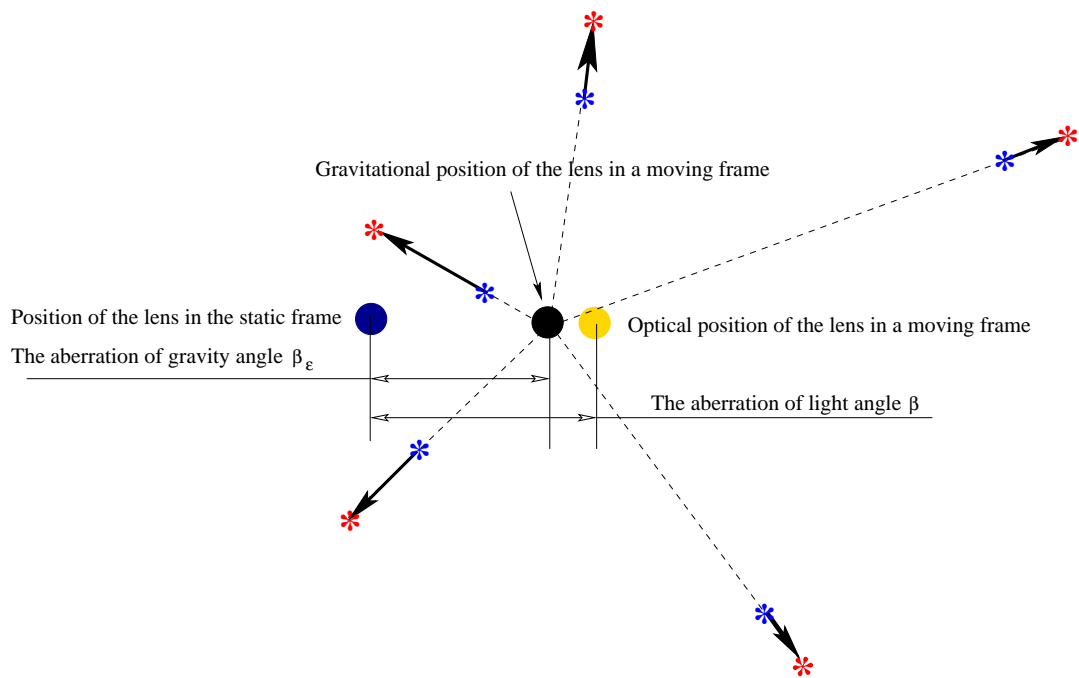}
\caption[Light deflection in the moving frame]{Deflection of light in a moving frame. The apparent optical
positions of the stars and the light-ray deflecting body are shifted by the aberration of
light from their static-frame positions. If the Lorentz invariance of
gravity is broken, the body's center of gravity,
determined from the gravitational deflection of light, will not
coincide with its optical position. Light-ray deflection
experiment conducted in a moving frame measures the difference
$\delta\beta=\beta_\epsilon-\beta$ defined in equation
(\ref{qk7c}). In general relativity $\delta\beta=0$.  }
\label{fig2}
\end{figure*}
\clearpage
\begin{figure*}
\includegraphics[height=21cm]{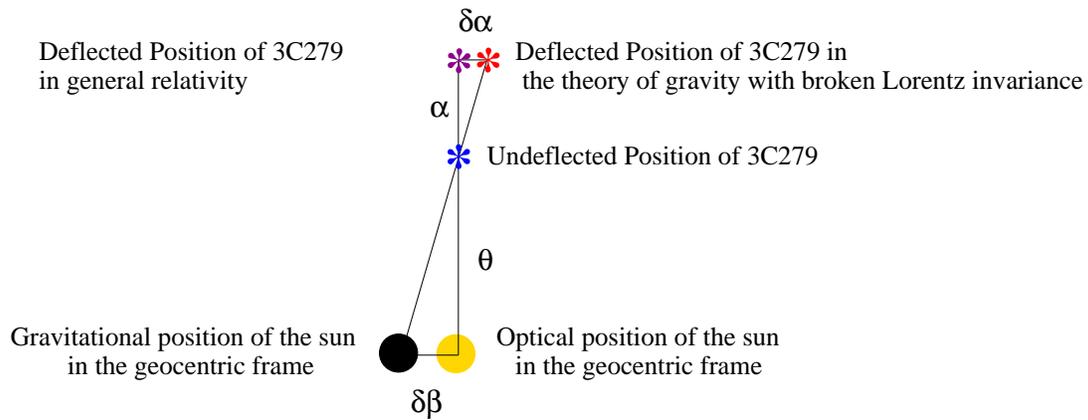}
\caption[Light deflection in the geocentric frame]{The difference
between the aberration angles for gravity force and light in
the plane of the sky in the geocentric frame. The angular separation
of 3C279 from the Sun is $\theta=d/(1 {\rm AU})$. The apparent position
of the quasar is shifted in the sky by the force of solar gravity at
the angle $\alpha$ if $c_{\rm g}=c$. In case of $c_{\rm g}\not= c$ the quasar is displaced by gravimagnetic force in the direction of orbital motion of the earth at a
small angle $\delta\alpha$ given by equation (\ref{ura}). It
corresponds to the angle $\delta\beta
=(\epsilon-1)(v_\odot/c)$ between the directions to the gravitational
and optical positions of the Sun. The relationship $\delta\alpha/\alpha=\delta\beta/\theta$ is
held.}
\label{fig3}
\end{figure*}
\clearpage
\normalsize
\begin{center}
\begin{tabular}{lccccccc}
\multicolumn{8}{c}{TABLE 1: ESTIMATED EXPERIMENTAL SENSITIVITY TO $\gamma$ and
ABERRATION } \\ \\\hline\hline \multicolumn{1}{c} {(1)} &
\multicolumn{1}{c} {(2)} & \multicolumn{1}{c} {(3)} &
\multicolumn{1}{c} {(4)} & \multicolumn{1}{c} {(5)} &
\multicolumn{1}{c} {(6)} & \multicolumn{1}{c} {(7)} &
\multicolumn{1}{c} {(8)} \\

\multicolumn{1}{c}  {Date} &
\multicolumn{1}{c}  {(Sol Rad)} &
\multicolumn{1}{c}  {(mas)} &
\multicolumn{1}{c}  {(mas)} &
\multicolumn{1}{c}  {(mas)} &
\multicolumn{1}{c}  {(mas)} &
\multicolumn{1}{c}  {($10^{-5}$)} &
\multicolumn{1}{c}  {($10^{-2}$)} \\

\hline 
October 1  &28.9 &  35 & 0.0 & 0.039 & 0.015 & -- & --- \\ 
October 5  &10.0 & 176 & 0.1 & 0.375 & 0.015 & 21 & 4.0 \\
October 6  & 6.3 & 279 & 0.3 & 0.945 & 0.015 & 12 & 1.5 \\ 
October 7  & 2.5 & 704 & 9.1 & 6.000 & 0.090 & 33 & 1.5 \\ 
October 9  & 5.2 & 338 & 2.1 & 1.386 & 0.025 & 13 & 1.8 \\ 
October 10 & 9.0 & 196 & 0.2 & 0.363 & 0.015 & 13 & 4.1 \\ 
October 11 &12.8 & 137 & 0.1 & 0.229 & 0.015 & 19 & 6.6 \\ 
October 18 &39.6 &  40 & 0.0 & 0.018 & 0.015 & -- & --- \\
\hline All days & & & & & & ~7 & ~ 1.0 \\ 
\hline\hline\\
\multicolumn{8} {l} {(1) = Mean observing time} \\ 
\multicolumn{8} {l} {(2) = Sun-3C279 separation} \\
\multicolumn{8} {l} {(3) = Gravitational bending of light} \\ 
\multicolumn{8} {l} {(4) = Estimate coronal bending at 23 GHz} \\ 
\multicolumn{8} {l} {(5) = Gravitational aberration of Sun} \\
\multicolumn{8} {l} {(6) = Expect positional sensitivity; 0.015 + 1\% of coronal bending} \\ 
\multicolumn{8} {l} {(7) = Accuracy of $\gamma-1$} \\
\multicolumn{8} {l} {(8) = Accuracy of aberration of gravity $\epsilon-1$} \\
\end{tabular}
\end{center}
\end{document}